\documentclass[acmsmall,review=false,screen,anonymous=false]{acmart}

\usepackage{amsmath}
\usepackage{amsfonts}
\usepackage{dsfont}
\usepackage{mathtools}
\usepackage{graphicx}
\usepackage{booktabs} 
\usepackage[inline]{enumitem}
\usepackage{balance}
\usepackage[skip=0pt]{caption}
\usepackage{multirow}
\usepackage{siunitx}
\usepackage{acronym}
\usepackage{subfig}
\usepackage{xcolor}
\usepackage{colortbl}
\usepackage{tabularx}
\usepackage{filecontents}
\usepackage{etoolbox,siunitx}
\usepackage{pgfplots}
\usepackage{caption}
\usepackage{algorithm}
\usepackage{algpseudocode}
\usepackage{xspace}
\usepackage{soul}
\usepackage{hyperref}

\acmJournal{TOIS}
\acmVolume{1}
\acmNumber{1}
\acmArticle{1}
\acmYear{2024}
\acmMonth{7}

\newcommand{\as}[1]{\textcolor{purple}{#1}}

\algrenewcommand\algorithmicrequire{\textbf{Input:}}
\algrenewcommand\algorithmicensure{\textbf{Output:}}

\newcommand{\RomanNumeralCaps}[1]
    {\MakeUppercase{\romannumeral #1}}
\AtBeginDocument{%
  \providecommand\BibTeX{{%
    \normalfont B\kern-0.5em{\scshape i\kern-0.25em b}\kern-0.8em\TeX}}}

\newcommand{\taskone}{Task \RomanNumeralCaps{1}\xspace}
\newcommand{\tasktwo}{Task \RomanNumeralCaps{2}\xspace}

\newcommand{\typeone}{Type \RomanNumeralCaps{1}\xspace}
\newcommand{\typetwo}{Type \RomanNumeralCaps{2}\xspace}

\newcommand{\itti}{Itti \& Koch\xspace}

\acrodef{OD}{outlier detection\xspace}
\acrodef{TTF}{time to find\xspace}
\acrodef{RT}{reaction time\xspace}
\acrodef{GBVS}{graph-based visual saliency\xspace}
\acrodef{AOI}{area of interest\xspace}
\acrodef{TTFF}{time to first fixation\xspace}
\author{%
Fatemeh Sarvi%
}
\orcid{0000-0002-6284-1636}
\affiliation{%
  \institution{
  Albert Heijn
  \city{Amsterdam}
  \country{The Netherlands}%
  }
}  
\email{fatemeh.sarvi@ah.nl}

\author{%
Mohammad Aliannejadi%
}
\orcid{0000-0002-9447-4172}
\affiliation{%
  \institution{
  University of Amsterdam
  \city{Amsterdam}
  \country{The Netherlands}%
  }
}  
\email{m.aliannejadi@uva.nl}

\author{%
Sebastian Schelter%
}
\orcid{0000-0003-4722-5840}
\affiliation{%
  \institution{
  BIFOLD \& TU Berlin
  \city{Berlin}
  \country{Germany}%
  }
}
\email{schelter@tu-berlin.de}

\author{%
Maarten de Rijke%
}
\orcid{0000-0002-1086-0202}
\affiliation{%
  \institution{
  University of Amsterdam
  \city{Amsterdam}
  \country{The Netherlands}%
  }  
}  
\email{m.derijke@uva.nl}

\def\authors{Fatemeh Sarvi, Mohammad Aliannejadi, Sebastian Schelter, and Maarten de Rijke}

\begin{document}


\title[]{Understanding Visual Saliency of Outlier Items in Product Search}

\renewcommand{\shortauthors}{Sarvi et al.}


\begin{abstract}
In two-sided marketplaces, items compete for user attention, which translates to revenue for suppliers. Item exposure, indicated by the amount of attention items receive in a ranking, can be influenced by factors like position bias. Recent work suggests that inter-item dependencies, such as \emph{outlier items} in a ranking, also affect item exposure. 
Outlier items are items that observably deviate from the other items in a ranked list w.r.t. task-specific, presentational features.
Understanding outlier items is crucial for determining an item's exposure distribution.

In our previous work, we investigated the impact of different presentational features on users' perception of outlierness in e-commerce search result pages. 
By modeling the problem as visual search tasks, we compared the observability of three main features: price, star rating, and discount tag. We found that participants perceive these features differently in terms of attention and reaction times. Various factors, such as visual complexity (e.g., shape, color), discriminative item features (e.g., a solitary discount tag), and value range, affect item outlierness.
These factors can be categorized into two main classes: \emph{bottom-up} and \emph{top-down}. Bottom-up factors are driven by visual properties such as color, contrast, and brightness, while top-down factors are influenced by cognitive processes such as expectations and prior knowledge.

In this extension of our previous work, we deepen our analysis of user perceptions of outliers. 
In particular, we focus on two key questions left unanswered by our previous work:
\begin{enumerate*}[label=(\roman*)]
    \item What is the effect of isolated bottom-up visual factors on item outlierness in product lists?
    \item How do top-down factors influence users' perception of item outlierness in a realistic online shopping scenario?
\end{enumerate*}

We start with bottom-up factors and employ visual saliency models to evaluate their ability to detect outlier items in product lists purely based on visual attributes. 
Then, to examine top-down factors, we conduct eye-tracking experiments on the same task as our previous visual search experiment: online shopping. This time, we design the task as a simulated e-commerce environment, mimicking a popular European online shopping platform to be more representative of real-world scenarios. Moreover, we employ eye-tracking to not only be closer to the real-world case but also to address the accuracy problem of reaction time in the visual search task. In our experiments, participants interact with realistic product lists, some containing outliers w.r.t.~different presentational features, such as image, price, and discount tag, at different positions. 

Our experiments show the ability of visual saliency models to detect bottom-up factors, consistently highlighting areas with strong visual contrasts and attention hotspots. 
While the well-known \itti model detects general visual attention patterns in an image, a \acl{GBVS} model (\acs{GBVS}) identifies visual anomalies more effectively. However, one should be cautious about the limitations of these models. Visual saliency models only rely on bottom-up factors, making them naive in that they do not distinguish between separate product features or compare them against each other.

The results of our eye-tracking experiment for lists without outliers show that despite being less visually attractive, product descriptions captured attention the fastest, indicating the importance of top-down factors and user knowledge of the task.  
Our observations in lists with visual outliers suggest that outliers and their immediate neighbors attracted attention faster (in terms of time to first fixation), which is in line with our findings from the visual search task. 
However, in our eye-tracking experiments, we observed that outlier items engaged users for longer durations (in terms of fixation count and time spent) compared to non-outlier items. This effect was consistent across different outlier features (image, price, discount tag) and various positions within the list. 
\end{abstract}

%
%



\maketitle

\acresetall


\section{Introduction}
\label{sec:intro}
 
In two-sided marketplaces, items compete for attention from users since attention translates to revenue for suppliers. 
Item exposure is an indication of how much attention each item receives from users. 
Effective estimation of item exposure is crucial for challenges such as item fairness~\citep{biega2018equity, mehrotra2018towards,  morik2020controlling, sapiezynski2019quantifying, singh2018fairness, singh2019policy, yadav2019fair, diaz2020evaluating} and bias in counterfactual learning to rank~\citep{joachims2005accurately, joachims2017unbiased,ovaisi2020correcting,agarwal2019addressing, vardasbi2020inverse,DBLP:conf/chiir/NovinM17}.
Various modeling assumptions have been proposed for item exposure estimation in ranking. 
Widely used modeling assumptions made to estimate item exposure include inter-item independence and definitions of exposure as a function of an item's position in a ranking. However, recent research has introduced different types of inter-item dependencies that influence exposure distribution on a ranked list, such as attractiveness bias~\citep{yue2010beyond}, context bias~\citep{wu2021unbiased} and outlier bias~\citep{sarvi2021understanding, sarvi2022bias}.

In this work, we focus on a phenomenon that accounts for a specific type of inter-item dependency~\citep{sarvi2021understanding,sarvi2022bias}: the existence of \emph{outlier items} in a ranked list may affect the exposure that all items in the list receive. 
Outlier items are those that observably deviate from the rest of the items in a ranked list w.r.t.\ task-specific, presentational features. 
presentational features are item features that are visible to users when examining the result page like the price of a product in product search. 

We have previously shown that the presence of outlier items may result in attention being distributed in a different way than on a list without outliers~\citep{sarvi2021understanding,sarvi2022bias}. For instance, on an e-commerce search result page, adding a red-colored discount tag as a discriminative feature to only one product can attract more attention to it irrespective of its position or relevance to the query, thereby deviating from the exposure distribution that is estimated only based on position-based assumptions. 

The perception and visual search communities have conducted many studies into how the human brain can immediately identify recognizable objects like outliers in an image and how different visual attributes (e.g., color, shape) can add to the complexity of this task~\citep{treisman1980feature,giovannangeli2022color,wolfe1998can,shen2003guidance}.
Since presentational features can be composed of multiple visual attributes, the relation between their use as discriminative features and their perception by users is complex.

\subsection{Presentational features and attention}
To gain a better understanding of the relationship between presentational features and user perception, we have previously compared different presentational features from the e-commerce search domain~\citep{sarvi2021understanding,sarvi2022bias}. We have provided insights into how different presentational e-commerce features impact users’ perception of the outlierness of an item on a search result page.
Informed by visual search studies, we have previously designed a set of crowdsourcing tasks where we compared the observability of different presentational features. The objective of these tasks is to find a target (i.e., outlier item) among distractors (i.e., non-outlier items), as fast as possible. Following previous work~\citep{treisman1980feature, duncan1989visual}, we used \acfi{RT} and accuracy in measuring the effort it takes to detect the target (outlier) among its distractors. We considered three observable features commonly used in e-commerce, viz.\ price, star rating, and discount tag. Previous work has shown the importance of these features in influencing users' purchase decisions~\citep{aggarwal2016font, kao2020effects}.

Our observations revealed that participants perceive different presentational features differently in terms of their attention and reaction times. Also, we found that the visual complexity of a feature can make it more observable to users. For example, a bright red background color of a discount tag makes it easier to spot than price tags that are shown as a number with regular font size and color.
These factors can be categorized into two main classes: \emph{bottom-up} and \emph{top-down}. Bottom-up factors are driven by visual properties such as color, contrast, and brightness, while top-down factors are influenced by cognitive processes such as expectations and prior knowledge.

In this extended study, to better understand the balance between what naturally grabs users' attention and what users prioritize during online shopping, we analyze these two types of factors separately.
In particular, we focus on two key questions left unanswered by our previous work:
\begin{enumerate*}[label=(\roman*)]
    \item What is the effect of isolated bottom-up visual factors on item outlierness in product lists?
    \item How do top-down factors influence users' perception of item outlierness in a realistic online shopping scenario?
\end{enumerate*}

\subsection{How different features contribute to the outlierness of an item}

We start with visual saliency models in Section~\ref{sec:visual-saliency} to examine the extent to which outlier products in a list can be detected based solely on bottom-up factors such as color, shape, and contrast. Next, in Section~\ref{sec:eyetracking-exp} we examine the effect of top-down factors by conducting eye-tracking experiments on the same task as our previous visual search experiment, i.e., online shopping. This time we design the task as a simulated e-commerce environment mimicking a popular European online shopping platform to be more representative of real-world scenarios.

Our experiments confirm that visual saliency models are effective in detecting bottom-up factors, consistently emphasizing areas of the image with high visual contrast. While the \itti model \citep{itti1998model} captures general patterns of visual attention in an image, the \acl{GBVS} model (\acs{GBVS}) \citep{harel2006graph} is better at identifying outlier regions of the image. However, it is important to acknowledge the limitations of these models. Visual saliency models depend solely on bottom-up factors, which means they cannot distinguish between different product features or assess them in relation to each other.

Our eye-tracking experiments suggest that product descriptions captured attention quickly, despite being less visually attractive, in lists without outliers. This finding indicates the importance of top-down features and other factors in play like center bias~\citep{tatler2007central, buswell1935people, foulsham2008can}. In lists with outliers, our analyses show that outliers and their immediate neighbors attracted attention faster and for longer durations compared to distant items. This effect was consistent across different outlier features (image, price, discount tag) and various positions within the list. We find that outlier items not only stand out among the list, but also receive more exposure as users spend more time examining them.

\subsection{Main contributions}
The main contributions of this work are:
\begin{enumerate}

    \item We build on our previous work that demonstrated how different presentational features (e.g., price, star rating, discount tags) impact user perception of outlierness in e-commerce search result pages, highlighting the key role of visual complexity in attention distribution;

    \item Through experiments with visual saliency models, we analyze the influence of bottom-up visual factors on item outlierness in product lists, confirming the effectiveness of the \acl{GBVS} model in detecting visual anomalies in ranked lists;

    \item Through eye-tracking experiments, we demonstrate the impact of top-down factors on user attention, showing that these factors can override bottom-up visual signals in online shopping scenarios;

    \item We show that outlier items and their close neighbors in ranked lists attract more attention and receive increased exposure (measured by engagement time), regardless of their position, due to their distinct observable features.
    
\end{enumerate}

\section{Background}
\subsection{Visual search}
Visual search has been a central approach in studying visual attention for many years. It allows researchers to turn everyday search activities, like finding a can opener in the kitchen, into controlled experiments that can be repeated in a lab setting~\citep{wolfe2010reaction}.

In a typical visual search task, individuals scan a visual field to locate a target object among other distracting items. Researchers examine how features such as color, shape, size, or orientation affect the speed and accuracy of finding the target~\citep{treisman1980feature}. One of the key theories explaining this process is the Feature Integration Theory (FIT) introduced by~\citet{treisman1980feature}. FIT proposes that visual search occurs in two stages: the pre-attentive stage and the focused attention stage. In the pre-attentive stage, basic features like color and shape are processed automatically and in parallel across the visual field. However, in the focused attention stage, when these features need to be combined to identify an object, the process becomes slower and requires more cognitive effort. This theory explains why finding a single, distinctive feature (such as a red dot among blue dots) is easier and faster than searching for an object that shares multiple features (like a red circle among red squares).
Another important concept in visual search is the difference between the stand-out effect and conjunction search, introduced by~\citet{wolfe1994guided}. When a target differs from all distractors by just one feature (such as color), it stands out, making the search quick and independent of the number of distractors. In contrast, when the target shares features with the distractors (for example, finding a red circle among red squares and green circles), the search becomes slower, and the task duration increases as the number of distractors grows.
Visual search tasks are typically assessed using \ac{RT} and accuracy, which help determine how quickly and efficiently a target can be identified among distractors~\citep{majeed2023anxiety,wolfe2020visual, palmer2011shapes, muller2003determining}.

\subsection{Visual saliency}

Visual saliency determines the perceptual selection of objects or regions that stand out and capture attention within a visual field~\citep{itti1998model}. It influences the control of visual attention, for example, in determining the next fixation points during visual exploration~\citep{veale2017visual,itti2007visual}.

Two types of factors influence the visual saliency of an object in a specific context: bottom-up and top-down factors~\citep{leiva2020understanding}. Bottom-up factors are primary visual attributes such as color, shape, size, and orientation. Objects that are unique with respect to such attributes tend to attain the observer's attention. For example, in an image mainly filled with green colors and shapes, the sudden appearance of a different color like red often makes people look at the red part~\citep{lu2012saliency,leiva2020understanding}. 
Top-down factors come from one's goals and what you expect to see~\citep{leiva2020understanding}. They are based on one's previous experience and their knowledge of the context. For example, when searching for something specific, like a red car, one is more likely to notice red cars first. 

Saliency effects can vary across different contexts and tasks. While some bottom-up factors like color contrast can be generalizable, top-down factors related to specific tasks can significantly influence what stands out. What is salient in one situation may not be in another. Researchers typically conduct specific experiments and use computational models to understand saliency within particular contexts and tasks rather than making broad generalizations~\citep{leiva2020understanding}.

\subsection{Outliers in ranking}
\label{sec:rw:outlier}
Outliers are data points that differ significantly from the rest of the data~\citep{wang2019progress}. They can represent unusual but important findings or potential errors. In any case, they are often seen as noise that can influence statistical analysis.
Various methods have been developed to detect outliers~\citep{scholkopf2001estimating, jin2006ranking, ramaswamy2000efficient, estimator1999fast} as it is essential to detect these anomalies in many research fields~\citep{li2020copod, wang2019progress}. However, the definition of outlier items can vary across domains~\citep{wang2019progress}. In this work, we adopt the definition of outliers in rankings from~\citet{sarvi2022understanding} who describe outliers as items that stand out based on observable features. observable features are visible characteristics that distinguish outlier items from their neighbors~\citep{sarvi2022understanding}. In this work, we create outlier items in ranking based on this definition and using different product properties as observable features.


\section{Preliminary Experiments: Visual Search}
\label{sec:pre-exp}
Studies in the field of visual search and cognitive science show that different visual attributes are processed differently by the brain~\citep{treisman1988features}. Inspired by these findings, and as the first step, we aim to verify if 
users notice deviations in different products' presentational features at different rates.
In this section, we describe the details of our crowdsourcing task which is formulated as a visual search experiment. 

\subsection{Crowdsourcing experiments}
We design our tasks as a visual search process~\citep{giovannangeli2022color, treisman1980feature, wolfe1998can,shen2003guidance}, where the objective is to find a target among distractors.
We focus on the domain of e-commerce search, where the distractors are non-outlier products in a ranking, and the target is the outlier products that differ in at least one presentational feature.
We compare three presentational features, namely, price, star rating, and discount tag.
When considering a discount tag, our task is close to a disjunctive search process known from visual search~\citep{treisman1980feature} that focuses on detecting a target that differs from the distractors in terms of a unique visual feature, e.g., the discount tag~\citep{mcelree1999temporal}.
When regarding price and star rating, our task is closer to conjunction search~\citep{treisman1980feature}, where the distractors exhibit at least one common feature with the target~\citep{shen2003guidance}. However, unlike conjunction search, in our task, the difference between the target and the distractors is in the values of the features, not the features themselves (e.g., the value of the product's price). 
In the rest of this paper, we refer to the target item as \emph{outlier}.

Following previous work~\citep{treisman1980feature, duncan1989visual}, we use \acfi{RT} and \emph{accuracy} to measure the effort it takes to detect the target (outlier) among its distractors.
The goal of this task is to examine and determine which presentational features are easier to detect by the workers, i.e., the shorter the \ac{RT} to find the outlier, the easier it is.

We perform our experiments using two tasks, where we build several synthetic product search result pages and examine how each feature contributes to the outlierness of an item, both separately and simultaneously. Below, we describe the different stages of our two tasks.

\subsubsection{Experimental design}
In the following, we describe the details of our experimental setup.
\paragraph{Page examination behavior}
We record several signals related to participants' page examination behavior and their interaction data, such as mouse hovering, scrolling, viewed items, clicks, and time spent on the task.  
To gain a more accurate estimate of \ac{RT}, we ask the participants to click on a \emph{Start} button after reading the instructions. The search result page appears only after the Start button has been clicked. We compute the task completion time from the moment the workers click the Start button.

\paragraph{Instructions}
In the instructions, we describe the overall goal of the research and the concept of an outlier in a search result page, providing tangible examples. We ask participants to scan and compare all items in a list and flag outliers as \emph{fast} as they can. We also ask them to fill out a questionnaire after completing the tasks. 

\paragraph{Participants}
We use Amazon Mechanical Turk as the platform for our crowdsourcing experiments, with workers based only in the U.S., with an approval rate of 95\% or greater. After quality control, we are left with 140 assignments (92 for \taskone and 48 for \tasktwo), submitted by 80 distinct participants. From the participants, $45\%$ are female,  $53\%$ male, and $2\%$ listed other genders. The majority of participants ($74\%$) are between 25 and 44 years old, with $5\%$ younger and $21\%$ older workers.

\paragraph{Post-task questionnaire}
We ask participants to fill out a questionnaire after completing the task. 
To gain more insight into workers' backgrounds and online shopping behavior, we instruct them to fill out questions on their demographics and familiarity with online shopping.
Moreover, to enable more effective results analysis, we ask the workers how much each product feature influences their everyday purchase decisions. 
To ensure that the workers understand the outlier definition, we ask them to answer a question about the definition of an outlier in search.

\paragraph{Quality control}
We follow three strategies for quality control. As part of the post-task questionnaire, we ask a multiple-choice question on the definition of outliers. All participants managed to answer this question correctly. Also, following~\citet{aniket2008crowdsourcing}, we ask workers to justify their choice in a few keywords. We only remove the responses of those participants who entered random tokens as justifications of their answers. 
We also remove the responses of those who revisit the instructions more than two times while performing the task, since response time is crucial in this study.

\subsubsection{Task I} 
In the first task, we evaluate how fast any of the three presentational features (price, star rating, discount tag) can be spotted on a search result page. To this end, we explicitly describe and mention the one feature at a time to the participants and ask them to scan the list and find up to two outlier items, \emph{only} with respect to the given feature. For instance, after providing a definition of outliers in the instruction, we mention that there are one or two outliers in terms of different values for price in the list and that they have to find them as fast as they can. We place one of the outliers at the top of the list and the other at the bottom. To avoid position and randomness bias, we keep the position of the outlier items fixed while other items are randomly placed in the list.

\citet{wolfe1998can} suggests that visual features including luminance, color, and orientation affect the \ac{RT} in a visual search task. 
Following this work and inspired by the experiments in~\citep{treisman1980feature}, we tested two variations of \taskone, namely \typeone and \typetwo, where we change the shape, color and value of the presentational features to study different magnitudes of deviation of the outliers from the rest.
In \typeone, we use features that more strongly discriminate between the outlier and the rest compared to \typetwo. For example, an outlier w.r.t.~price can be 10 or 2 times more expensive than other products. We use the former in \typeone and the latter in \typetwo. 
The same goes for star rating. 
Regarding the discount tag feature, we use the suggestions by \citet{wolfe1998can} to distinguish between the outlier items of \typeone and \typetwo.
In \typeone we use a bright red color as background with a bold white font stating that there is a special deal on the product, whereas, in \typetwo, we use a light green text without any background stating a $10\%$ discount.

\subsubsection{Task II}
Unlike \taskone, here we aim to examine the relative \ac{RT} for the three features (price, star rating, discount tag) when presented to the users simultaneously. 
To better compare the three observable features, we jointly present the different combinations of these features and analyze the behavior of the users.
While describing the three target features in this task, we do not mention to participants which features are being examined. Therefore, the workers are supposed to go through the list, examine all items with respect to all the features used in presenting the results, and then decide which items are outliers. 
Note that there are more than three features used to describe each item, for example, we used image, title and delivery information next to the price, discount tag and star rating.
Moreover, we indicate that the workers have to mark a maximum of three items as outliers. Here, we also randomize the position of the outlier items while making sure that they appear both at the top and bottom of the list. We run the task for all combinations of at least two of the three features.

\subsection{Results}
\label{sec:pre-exp:results}
In this section, we present the results of our crowd-sourcing tasks in terms of the performance and behavior of the workers under different experimental conditions.


\begin{table}[h]
   \vspace{-1mm}
   \setlength{\tabcolsep}{3.2pt}
      \caption{
         Worker performance metrics in terms of \ac{RT} and accuracy. Average (Avg.) and median (Med.) \ac{RT} in seconds is reported for the first and second outlier (out.~1 \& out.~2).
         }
      \label{table:task1}
      \centering
      \begin{tabular}{ll rr c rr r c}
        \toprule
       \multirow{2}{*}{\textbf{Type}}  &  &  \multicolumn{2}{c}{\textbf{\ac{RT} out.~1}} & & \multicolumn{2}{c}{\textbf{\ac{RT} out.~2}} &  
       \multirow{2}{*}{\textbf{Acc.}} \\ 
       \cmidrule{3-4} \cmidrule{6-7}
       & & Avg. & Med. &&  Avg. & Med. & & \\
         \midrule
         \multirow{3}{*}{\rotatebox[origin=c]{90} {Type 1}}  &
         Disc. tag & 4.22 & 3.62 && 8.90& 8.00 & 0.98 \\
         &Star rating & 4.81 & 3.41 && 9.67 & 8.14 & 0.97 \\
         &Price & 8.38 & 5.50 && 12.44 & 11.77 & 1.00 \\
        \midrule
         \multirow{3}{*}{\rotatebox[origin=c]{90} {Type 2}}  &
         Disc. tag & 19.84 & 17.88 && 25.57 & 26.88 & 0.99 \\
         &Star rating & 10.96 &8.11 && 17.36 & 12.42 & 0.99\\
         &Price & 10.62  & 6.03 && 14.21 & 11.90 &  0.98\\
         \bottomrule
      \end{tabular}
   \end{table}
   
\noindent%
%
   \begin{figure}[h]
   \vspace*{-4mm}
       \subfloat[]{%
             \includegraphics[width=0.5\columnwidth]{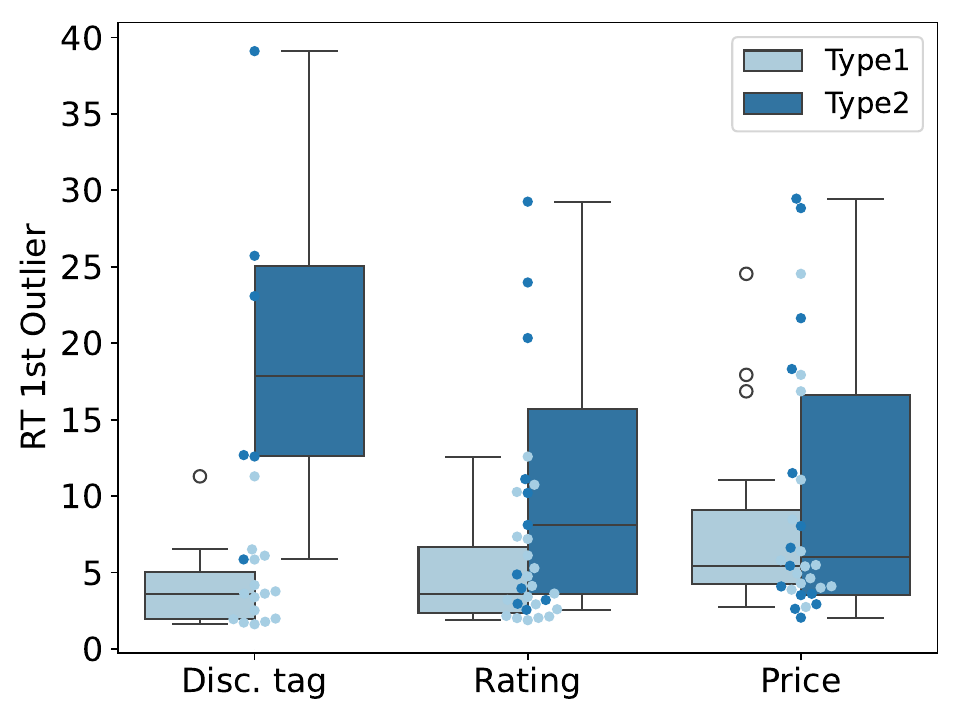}
             \label{fig:boxplot-out1}}
       \subfloat[]{%
             \includegraphics[width=0.5\columnwidth]{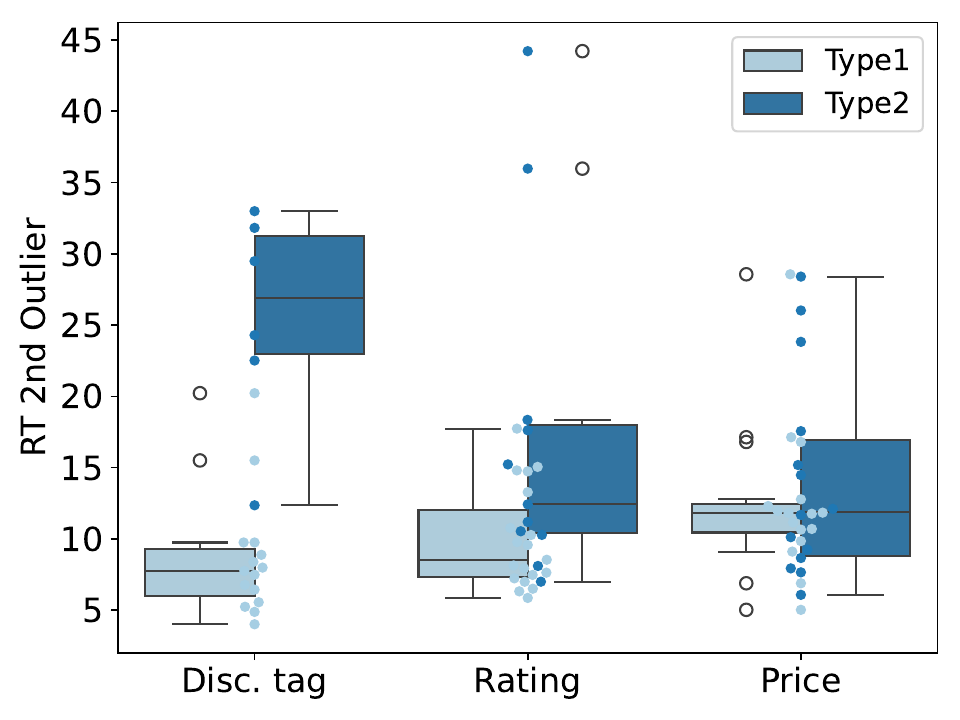}
             \label{fig:boxplot-out2}}
   \caption{Distribution of the \ac{RT} for the (a) first and (b) second outliers in both variations of \taskone.}
   \label{fig:time-boxplot}
   \vspace*{-1mm}   
   \end{figure}
%

\subsubsection{Reaction time and accuracy}
Following~\citep{treisman1980feature, duncan1989visual}, we use \acf{RT} and
accuracy to measure the effort it takes to detect the outlier
among non-outlier distractors.
Table~\ref{table:task1} summarizes participants' average responses to \taskone in terms of \ac{RT} and accuracy.
Accuracy is high for all variations of \taskone with a maximum of $1.00$ for price in \typeone and a minimum of $0.97$ for star rating.
We conclude from the high accuracy values that the workers grasped the concept of outlier in a ranked list and were able to accurately find them in the list.
Next, we compare the time that the workers take to spot the outliers.  
Table~\ref{table:task1} shows that for \typeone outliers  participants spotted the discount tags faster than the other two features in \taskone. This is followed by star rating with a slightly higher recorded \ac{RT}. As expected, we see that on average it took participants almost twice as long to find the price outliers.
We conducted a one-way ANOVA test on \ac{RT} for first outliers in \typeone. Results show that the differences are statistically significant with $p$-value $ < 0.05$.

Detecting discount tags is similar to a disjunctive visual search process, which has been shown to be easier to solve compared to conjunctive search (i.e., star rating and price) \cite{treisman1980feature}. Moreover, users favor simple options when they act under time pressure \cite{DBLP:conf/chiir/Azzopardi21}, which can lead to being biased towards easy-to-detect visual features such as discount tag.
The higher \ac{RT} for price can be attributed to the fact that certain visual features, including color and shape, are processed early in the brain using pre-attentive processes~\citep{treisman1980feature}.
Star rating and discount tag have more visual characteristics regarding shapes and colors, however price is more simply presented in the product descriptions.

Another related aspect is the unknown range of the price values.
This is less crucial for star rating or discount tag since the former has a range between 1 to 5 and latter is a binary feature.

\subsubsection{\typeone vs.~\typetwo}
Next, we compare the results of \typeone and \typetwo outliers. Our goal is to understand how much changing the magnitude of the deviations in terms of different features affect user performance. 
One can compare different ranges of deviations on specific features to model the relationship between the the deviations and user performance, but we leave this as future work and only compare two variations.
The results in the upper and lower parts of Table~\ref{table:task1} suggest that the reduced magnitude of deviations in all features leads to higher \ac{RT}.
\citet{duncan1989visual} study the same effect by pointing out that when outlier to non-outlier similarity increases, the task becomes more difficult. 
Similarly, 
we see that \ac{RT} increases for all the features, and for both the first and second outliers.

Moreover, we see in Figure~\ref{fig:time-boxplot} how the \ac{RT} distribution of the two outlier variants differ for \taskone.
As expected, the plots show a higher \ac{RT} for all features, and both outliers. However, it is interesting to note that we observe the lowest relative effect on the price ($26.73\%$ increase), compared to star rating ($127.86\%$) and discount tag ($370.14\%$). 
We relate this to the visual nature of discount tag and star ratings. Reducing the color contrast of discount tag would have a greater impact on the user's ability to detect it among the distractors, compared to a different price ranges as the user still has to carefully check the prices to detect the outlier.
Regarding the accuracy we see no drop, suggesting that even a more subtle deviation in observable feature can be detected by many users.

\subsubsection{Feature combinations}
\begin{figure}   
  \vspace{-3mm}         
  \subfloat[]{%
        \includegraphics[width=0.5\columnwidth]{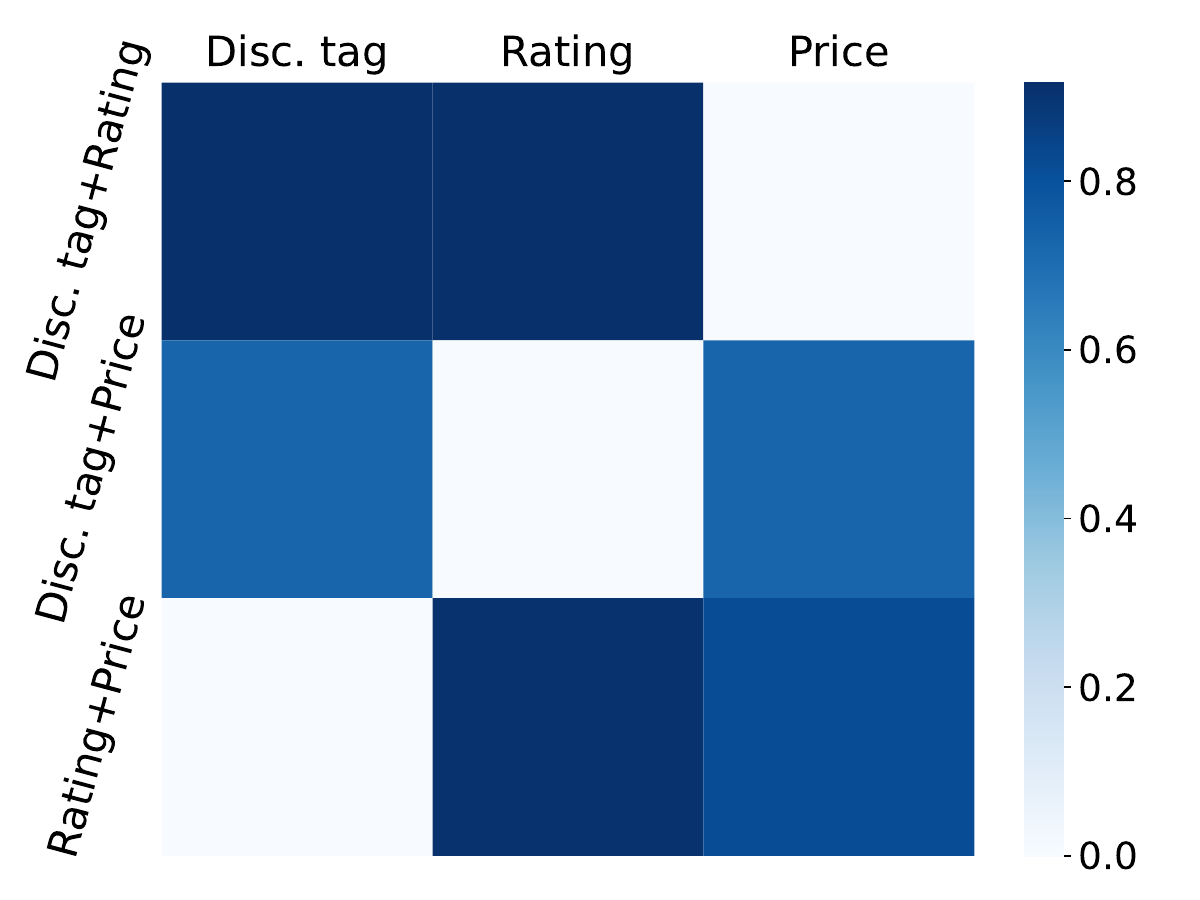}
        \label{fig:recall-heatmap}}
  \subfloat[]{%
        \includegraphics[width=0.495\columnwidth]{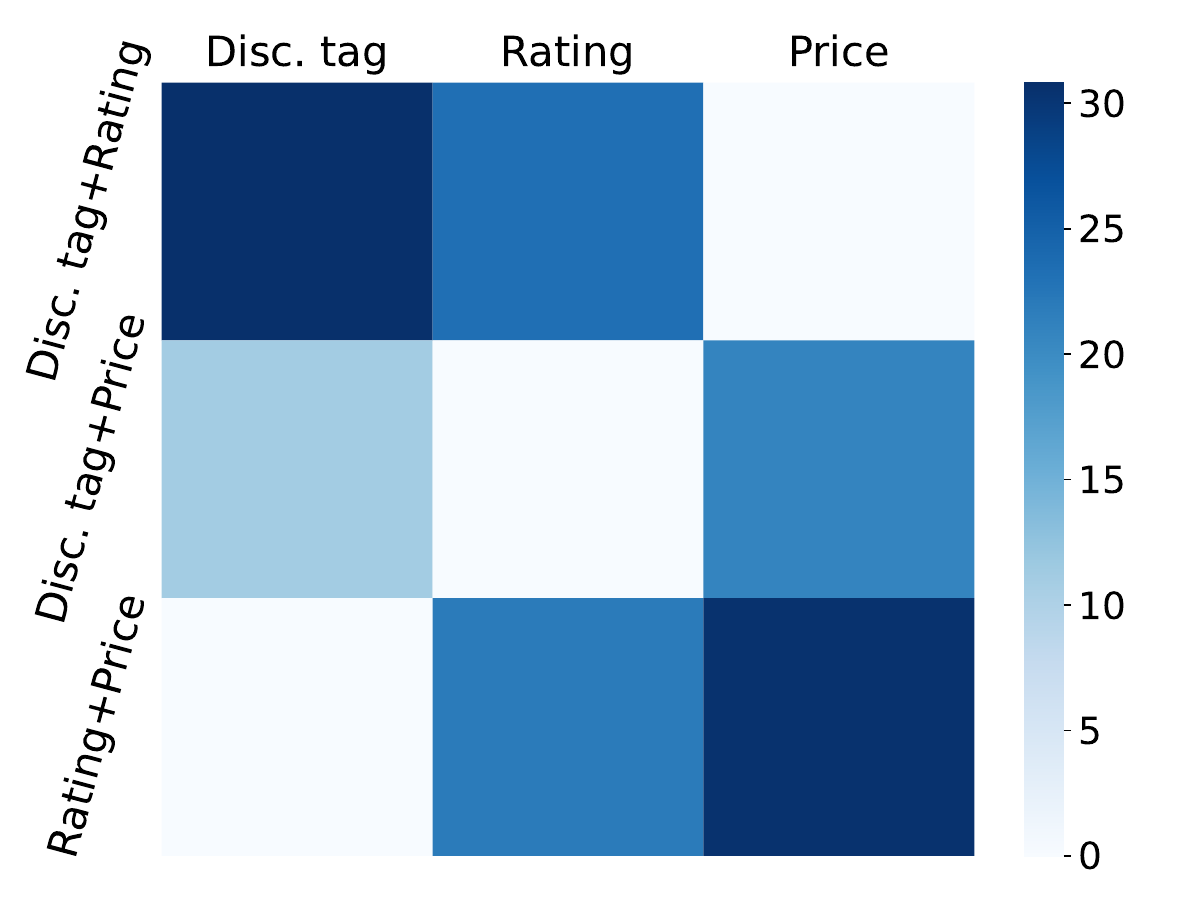}
        \label{fig:time-heatmap}} 
  \caption{(a) Recall and (b) \ac{RT} for combinations of observable features. Y-axis shows the metric of the corresponding feature and x-axis shows the second feature used in the combination.}   
\end{figure}

Figure~\ref{fig:recall-heatmap} shows recall values for combinations of features, where the y-axis indicates recall of a combination of features and the x-axis indicates the value for a specific feature. 
As expected, detecting the outlier w.r.t.~price is more difficult for participants (on average, $1.3\%$ and $7.3\%$ lower values than for discount tag and star rating).
In terms of \ac{RT}, Figure~\ref{fig:time-heatmap} confirms our findings in Table~\ref{table:task1}, except for the combination of discount tag and star rating, where on average participants found star rating ${\sim}7.5$ seconds faster than discount tag. Perhaps, it is because the average position of the outlier w.r.t.~discount tag is lower than star rating ($12$ and $9.5$, respectively). 


\section{Extended Experiments: Visual Saliency and User Attention}
\label{sec:eye-tracking-exp}

In the previous section, we investigated how individuals perceive and react to different product features (whether outliers or not) within a list through controlled visual search tasks. We measured participants' \ac{RT} and accuracy as they detect outliers among regular items in ranked lists.
The findings revealed that various factors, such as visual complexity (e.g., shape, color), discriminative item features (e.g., a solitary discount tag), and value range, affect user perception of item outlierness.
These factors can be categorized in two main classes: bottom-up and top-down. 
To better understand the balance between what naturally grabs users' attention and what users prioritize during online shopping, in this extended study, we analyze these two types of factors separately. We aim to answer the following research questions:

\begin{enumerate}[label=RQ\arabic*,leftmargin=*]
    \item \label{rq:saliency_models} What is the effect of
isolated bottom-up visual factors on item outlierness in product lists?
    \item \label{rq:eye-tracking} How do top-down factors influence users’ perception of item outlierness in a realistic online shopping scenario?
\end{enumerate}
We start with visual saliency models in Section~\ref{sec:visual-saliency} and examine the extent to which outlier products in a list can be detected based solely on bottom-up factors such as color, shape, and contrast. Next, in Section~\ref{sec:eyetracking-exp} we examine the effect of top-down factors by conducting eye-tracking experiments on the same task as our previous visual search experiment, i.e., online shopping. This time we design the task as a simulated e-commerce environment mimicking a popular European online shopping platform to be more representative of real-world scenarios.
Formulating the task as a visual search experiment is valuable for gaining initial insights into user behavior and attention patterns, however, it has two main limitations that we aim to address in our new setup:

\paragraph{Representativeness of e-commerce scenarios} 
Visual search tasks are structured and goal-driven, where participants are explicitly instructed to fulfill a specific goal (in our case: find the outlier item as quickly as possible). However, in real-world online shopping, users engage in more open-ended exploration, distributing their attention across multiple items and features without a clear target in mind. This contrast in task characteristics can lead to variations in user behavior, making it challenging to generalize findings from controlled visual search experiments to the real-world shopping case.
In this section, we employ a more realistic experimental design that mimics a popular e-commerce platform in Europe. 

\paragraph{Accuracy of \ac{RT}}

Much previous work uses \ac{RT} as a proxy to estimate user attention in visual search tasks~\citep{majeed2023anxiety,wolfe2020visual, palmer2011shapes, muller2003determining}, however, it might not always be accurate because it relies on participants' interpretations of our instruction. While we instructed participants to find the outlier as fast as they could, it is not straightforward to ensure they consistently follow this instruction. Response times can be influenced by factors beyond attention, or individual differences, such as external environmental distractors or lack of focus and motivation. Therefore, \ac{RT} alone may not provide a precise measure of attention.

In this extended study, we use eye-tracking as the tool for capturing and analyzing user attention. Attention, by its nature, is an internal and subjective experience~\citep{shi2013information,meyer2007understanding,chocarro2022attention}. Declarative methods, like surveys or self-reports, are inadequate when used alone for measuring user attention~\citep{ariely2010neuromarketing}. 
Eye-tracking provides objective and direct measurements of attention, making it a more reliable choice for understanding how users engage with visual stimuli~\citep{ariely2010neuromarketing,chocarro2022attention}.

\subsection{Visual saliency maps}
\label{sec:visual-saliency}
Visual saliency determines the regions that stand out and capture attention within a visual field~\citep{itti1998model}. In our task of detecting outlier product feature in search result list, it allows us to predict which product features are expected to attract more attention due to their visual properties, such as color contrast, size, or pattern complexity.
Visual saliency can serve as a benchmark or baseline for comparing the inherent attention-grabbing properties of different product features. It can also detect observable outliers purely based on bottom-up factors. In this section we aim to answer~\ref{rq:saliency_models}.
In the rest of this section we describe the models used for this experiment and our stimuli design.

\paragraph{Models}
Visual saliency models create density maps that depict the extent to which individual pixels grab attention relative to others (see Figure~\ref{fig:visual-sal-maps-list}). These models can be classified into bottom-up and data-driven. Bottom-up models are based on primary visual attributes such as shape and color, while data-driven models are trained using eye movement data together with some architectural assumptions inspired by bottom-up models~\citep{leiva2020understanding}. 
To answer~\ref{rq:saliency_models} we use two well-known bottom-up models that are often used as baseline models in the literature: \itti~\citep{itti1998model} model and \ac{GBVS}~\citep{harel2006graph} model. \citet{itti1998model} proposed a saliency-based visual attention model to extract visual features as computed via linear center-surrounded operations with Gaussian pyramids for intensity, color, and orientation. The \ac{GBVS} model is an extension of the \itti model that forms graph-based activation maps from visual features and normalizes them to highlight conspicuity. The global visual feature extraction and graph-based activation maps enable the model to capture saliency maps at the global level.

\paragraph{Stimuli description}
To answer~\ref{rq:saliency_models} we captured high-resolution screenshots of search result pages of Bol.com, a popular European e-commerce platform. Each screenshot is preprocessed to fit the input requirements of the saliency models. We then generated saliency maps using both models to highlight areas predicted to attract the most visual attention. 

For the analysis, we selected a subset of variants of product lists that include different product categories, different outlier positions and outlier types:
\begin{enumerate}
    \item a list of smartphones with an outlier image at position 3 (see Figure~\ref{fig:visual-sal-maps-mobile-list}),
    \item a list of monitors with an outlier discount tag at position 8 (see Figure~\ref{fig:visual-sal-maps-monitor-list}), and
    \item a list of office chairs with an outlier price at position 13 (see Figure~\ref{fig:visual-sal-maps-chair-list}).
\end{enumerate}
We generate the visual saliency maps for all these variants, illustrating both the full list (see Figure~\ref{fig:visual-sal-maps-list}) and a focused view around the outlier and its close neighboring items (see Appendix, Figure~\ref{appendix-fig:visual-sal-maps-list}). 

\subsection{Eye-tracking experiments}
\label{sec:eyetracking-exp}
In this section, we describe our experimental design to answer~\ref{rq:eye-tracking} through eye-tracking. We detail our methodology, experimental designs, and specific study goals.

\subsubsection{Experimental design}
In the following, we outline the key aspects of our experimental design.

\paragraph{Online shopping experience}
There are two critical factors influencing a customer's shopping experience: their goal or specific task, and product category~\citep{chocarro2022attention}.

Customer's goals refer to the different stages of the purchasing process, such as gathering information about products, comparing different options, and understanding delivery choices~\citep{tupikovskaja2021eye}.
In this study we focus on the \emph{Choice Task} as described by~\citet{chocarro2022attention}: ``Visit the website and select from those offered the product that most appeals to you based on the information provided.''

In addition, the category of the product has been recognized as a significant moderator in e-commerce. \citet{nelson1970information} divides product categories into two classes: search products and experience products.
\textit{Search products} are items that consumers can determine most of their attributes before purchasing. On the other hand, 
most attributes of \textit{experience products} are unknown to consumers before the purchase or the consumption.
In short, consumers can evaluate \textit{search} products by their features, brand, or price, while \textit{experience} items need senses for their evaluation~\citep{nelson1970information, chiang2003factors, weathers2007effects}. 

Previous studies suggest that user attention patterns can be different for different product categories~\citep{lee2021product, luan2016search, wang2014eye}. 
Therefore, following previous work~\citep{chocarro2022attention, huang2014we,kim2008effects,levin2003product,luan2016search}, we select 5 product categories with search and experience attributes:
\begin{itemize}
    \item \textit{experience} attribute: backpacks, office chairs, running shoes; and
    \item \textit{search} attribute: mobile phones, monitors.     
\end{itemize}
We use the backpacks category also for the calibration stage and the rest for actual analysis.

\paragraph{Stimuli description}
The stimuli in this study consist of product search result pages, mimicked after Bol.com (see Figure~\ref{fig:visual-sal-maps-list}).
Each page contains 15 distinct products. These products are characterized by various features, including product images, titles, descriptions, star ratings, review counts, pricing information, and, in some cases, discount tags. To reduce the complexity of the page we removed information about delivery, seller, and offer from the product features.

The products on our simulated search result pages are real items from Bol.com, chosen to match our study queries. However, they may not be exactly the same as the platform's search results. We made controlled modifications to certain product features for the purpose of our research. For instance, we may have adjusted item prices or added discount tags to items to study their effects. We provide the descriptions of these decisions where they were applied.

\paragraph{Online eye-tracking}
We use the \url{RealEye.io} online platform to run webcam-based eye-tracking experiments. \url{RealEye.io} eye-tracking is based on WebGazer, an eye tracking JavaScript library~\citep{papoutsaki2016webgazer}. 
Webcam-based eye-tracking has become popular in the eye-tracking community due to its ability to capture eye movements in real-world settings, relatively low cost, and high speed of data acquisition~\citep{wisiecka2022comparison}. Several recent studies use webcam-based eye-tracking for their perception and cognitive experiments~\citep{sarvi2021understanding,fazio2020consumer, federico2019tool, murali2021conducting, haldar2024learning, brandl2024evaluating}.
\citet{wisiecka2022comparison} report that they were able to obtain comparable results from \url{RealEye.io} compared to a lab experiment in tasks involving fixation (location-based metrics).
The average accuracy for individuals is reported as 113px; however, it is expected that the average error goes to zero in aggregated analysis with several participants.\footnote{\url{https://support.realeye.io/realeye-accuracy}}

\paragraph{Metrics and data collection}
In this study, \ac{AOI} is an analytical tool that provides eye movement metrics for user-defined areas of an image. We defined three \ac{AOI}s per product in the list that covers a product's image (on the left), the product's description (in the middle) and the product's price information (on the right) including discount tags if any.
We consider four eye-tracking measures to report our results based on participants’ eye fixations and for any \ac{AOI}~\citep{fiedler2020guideline}:
\begin{enumerate}
    \item fixation count: the number of fixations within an \ac{AOI}; more fixation means more visual attention; 
    \item time spent: shows the amount of time that participants spent on average looking at an \ac{AOI};
    \item \ac{TTFF}: the amount of time that it takes participants on average to look at the \ac{AOI} for the first time; and
    \item revisit count: indicates how many times participants looked back at the \ac{AOI} on average.
\end{enumerate}
To calculate these metrics, we aggregated eye movements on each \ac{AOI} on the list.

\paragraph{Instructions}
Participants were instructed to interact with the presented product lists as they typically would. While participants were not obligated to click on any items, they were encouraged to explore the entire page thoroughly. It was emphasized that we will ask questions regarding the content of each page after their exploration and we only accept submissions with reasonable answers to the questionnaires. 
We had 6 product lists each corresponding to a unique query, and participants had 90 seconds for exploring each single page. The first list was for calibration, so that the participants get familiar with the format of the experiment, therefore, we recorded the results only for the next 5 lists.

\paragraph{Post-task questionnaire}
Our questionnaire consisted of two sets. The first set was presented after each page was shown, focusing on the page's content, the products participants observed, their purchase recommendations, and whether they noticed anything unusual or intriguing.
The second set of questions was presented upon completion of the task and covered participants' demographics and online shopping habits. 

\paragraph{Procedure and implementation}
Participants were recruited through the Prolific platform\footnote{\url{https://prolific.io}} and, after receiving task instructions, were directed to RealEye for the eye-tracking experiments and questionnaires.

The participants were randomly exposed to different sets of lists.
Each set contains the same queries and products, but with modifications to place outliers in different positions. We also applied randomization within the set, to present different orders of queries to different participants.
The randomization of each set presented to participants was systematically managed through our backend logic, while the randomization of the order of lists within the sets is handled by the RealEye platform.

At the start of the task, each participant would initiate their session with a warm-up step, during which they explored a list of products. The results of this warm-up step are excluded from our subsequent analysis. 
Upon completion of all sessions and questionnaires, participants received a unique code, allowing Prolific to track their submissions.

\paragraph{Participants and procedure}
To ensure the participants' familiarity with the e-commerce platform, we only hired participants from the Netherlands and Belgium, where our e-commerce platform is active. To ensure data quality, we set a prerequisite that workers must have an approval rate of $95\%$ or greater. After quality control procedures, we are left with a total of 118 distinct participants.
$54.2\%$ of participants identified as female, $45\%$ as male, and $0.8\%$ selected other genders. In terms of age distribution, the majority of participants ($67.8\%$) were aged $18$ to $34$, $30\%$ were $35$ to $54$ years old, and the remainder were older than $54$.
All participants in our study used desktop computers with a webcam, ensuring a standardized viewing experience among all participants.

\subsubsection{Task I}
\paragraph{Overview}
In the first step of our eye-tracking experiment, we investigate how users examine product search result lists and understand the attention dynamics related to different product features. Specifically, we explore which features are more engaging in terms of time spent and fixation count and which ones capture attention faster in terms of \ac{TTFF}. While there are studies examining factors that influence users' viewing behavior on search results pages~\citep{lewandowski2021factors}, to the best of our knowledge, no previous research has focused on e-commerce.

\paragraph{Stimuli description}
As mentioned in Section~\ref{sec:eyetracking-exp}, the product lists used in our study were harvested directly from Bol.com. Each list contained 15 distinct products. To maintain consistency for this step, we ensured that no outlier products were present in the lists. To this end, we identified and replaced products that could potentially be perceived as outliers (see Section~\ref{sec:rw:outlier}). Several factors were considered during this process, including product images (content, color, background color), prices, discount tags, and user ratings.

Additionally, slight adjustments to item prices are applied in some cases to ensure they match the pricing patterns found in product lists. We used ``backpacks'' and ``running shoes'' as the search queries for this experiment. Participants were randomly assigned to one of these lists.

\subsubsection{Task II}
\paragraph{Overview}
In our second eye-tracking study, we aim to investigate how outlier product features influence the observability of products in search result lists. More specifically, we focus on the stand-out effect of different product features when presented as outliers. We aim to answer~\ref{rq:eye-tracking} by exploring how the presence of an outlier product feature affects the overall attention distribution among the list and how these effects differ among various product features.

\paragraph{Stimuli description}
We selected three product features for examination: price, image, and discount tag. 

To maintain an unbiased experimental design, we ensured that product category, position, and relevance do not introduce any bias into the results. To achieve this, we created unique combinations of product queries and product features, each featuring an outlier placed at positions \as{3, 8, or 13} within the product list. 

Each participant viewed each product query only once during the study, ensuring diversity and preventing familiarity from affecting their attention patterns. Furthermore, participants received lists with outliers w.r.t.~each product feature only once to prevent repetition bias and learning effects. The position of the outlier within the list was randomly assigned from the available options to address position bias.

\subsection{Results}
\subsubsection{Visual saliency maps}
In this section, we present our observations to answer~\ref{rq:saliency_models} using the full list view, however, the patterns are similar for the focused view images (see Figure~\ref{appendix-fig:visual-sal-maps-list}).
\begin{figure}[]
    \subfloat[]{%
    \includegraphics[width=0.26\columnwidth, height=5.2cm]{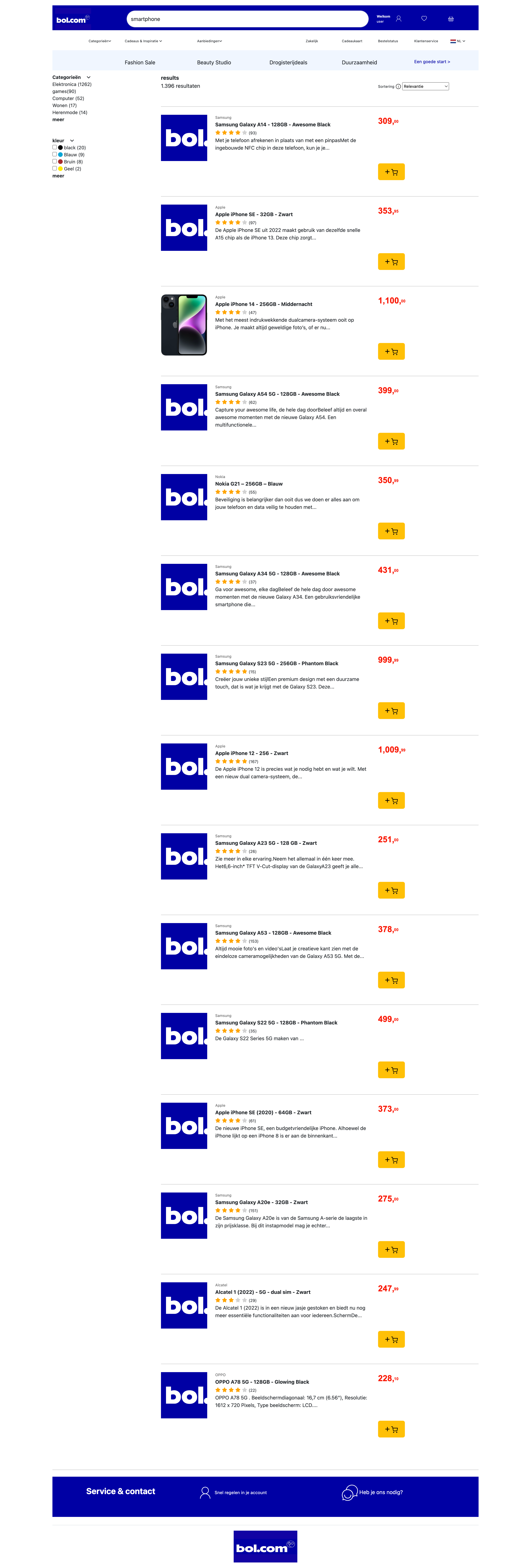}
    \label{fig:visual-sal-maps-mobile-list}
    }
    \subfloat[]{%
    \includegraphics[width=0.26\columnwidth, height=5.2cm]{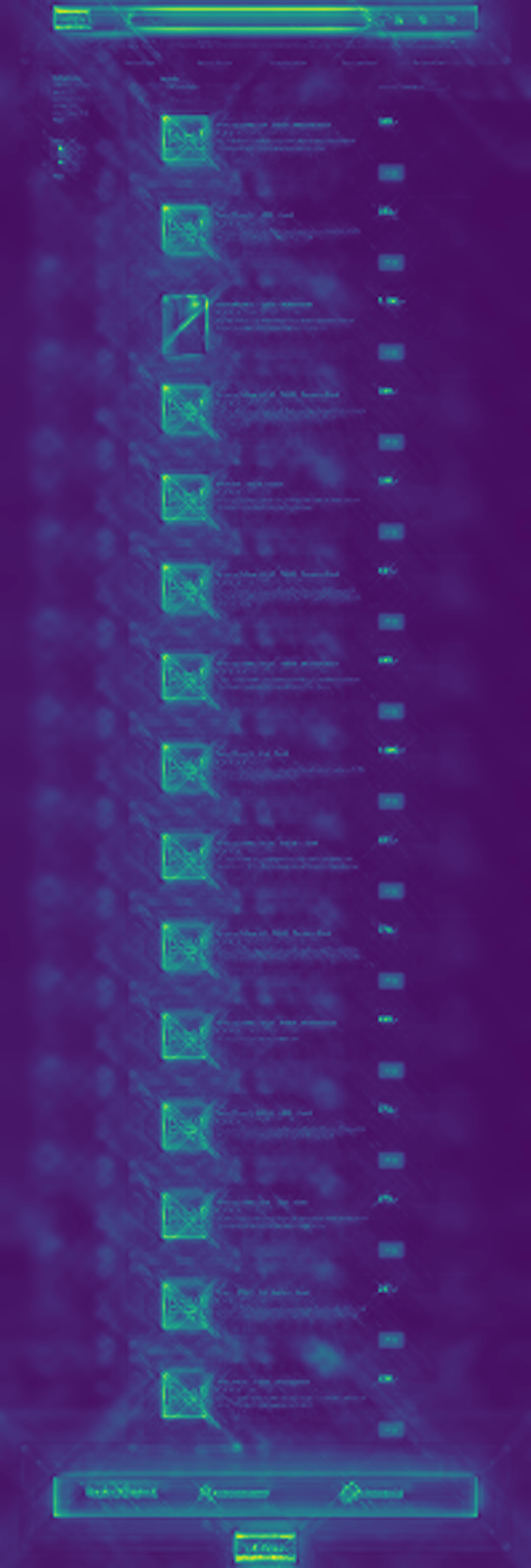}
    \label{fig:visual-sal-maps-mobile-list-itti}
    }
    \subfloat[]{%
    \includegraphics[width=0.26\columnwidth, height=5.2cm]{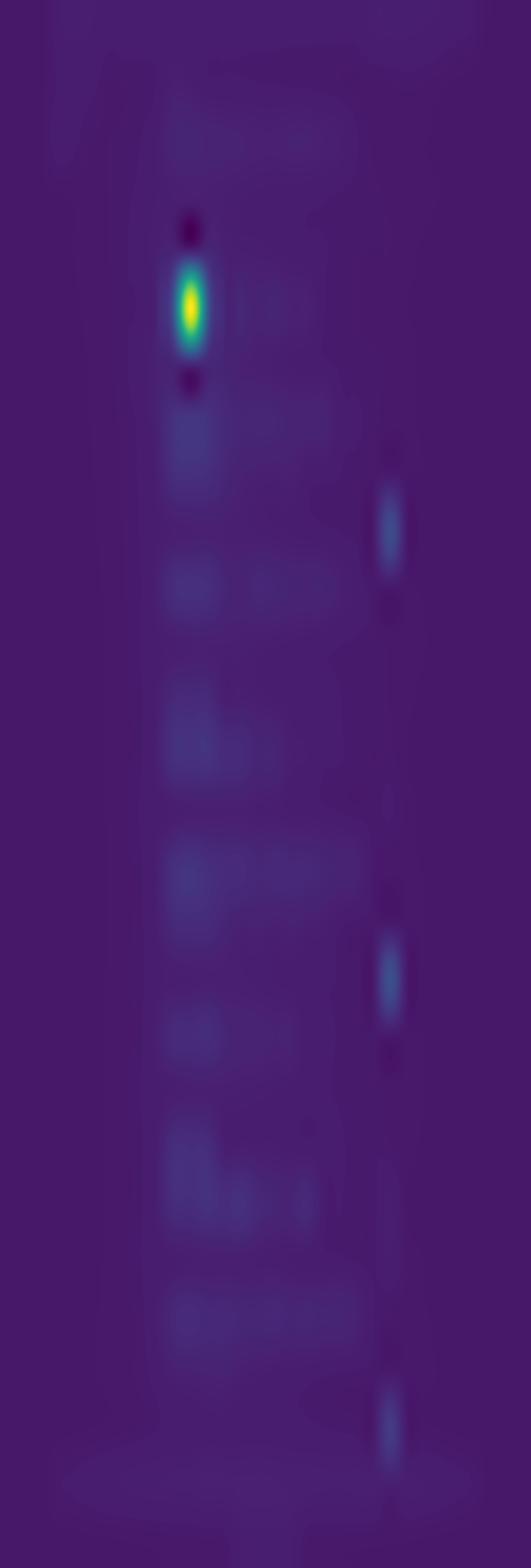}
    \label{fig:visual-sal-maps-mobile-list-gbvs}
    }

    \subfloat[]{%
    \includegraphics[width=0.26\columnwidth, height=5.2cm]{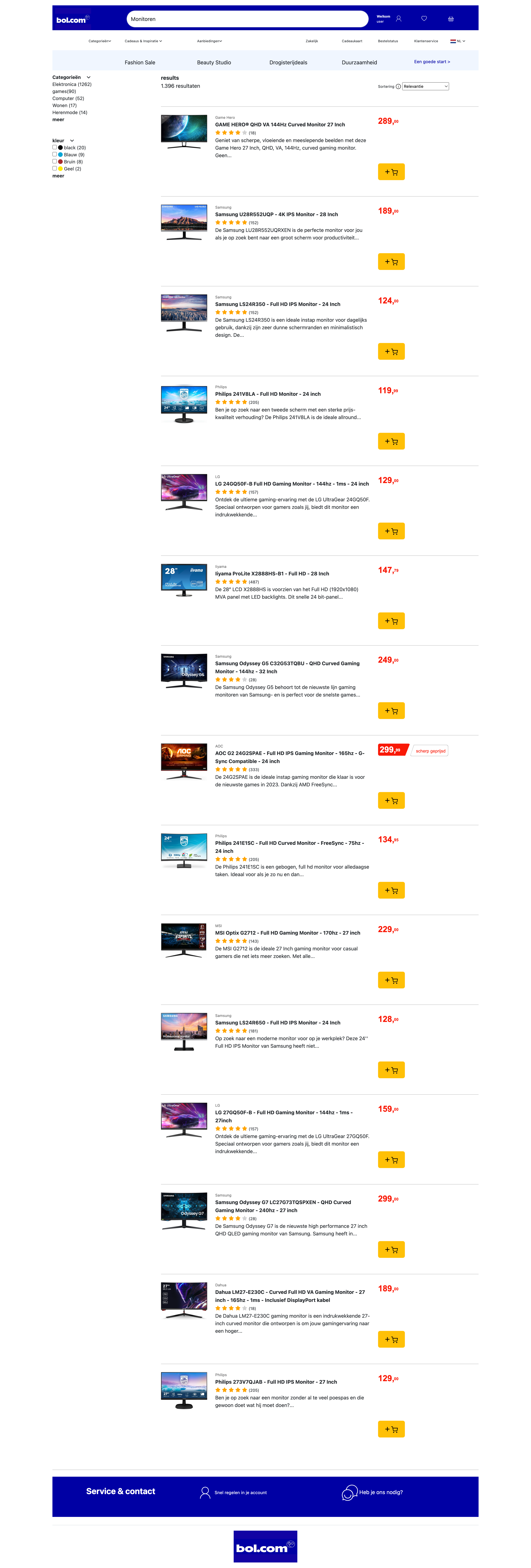}
    \label{fig:visual-sal-maps-monitor-list}
    }
    \subfloat[]{%
    \includegraphics[width=0.26\columnwidth, height=5.2cm]{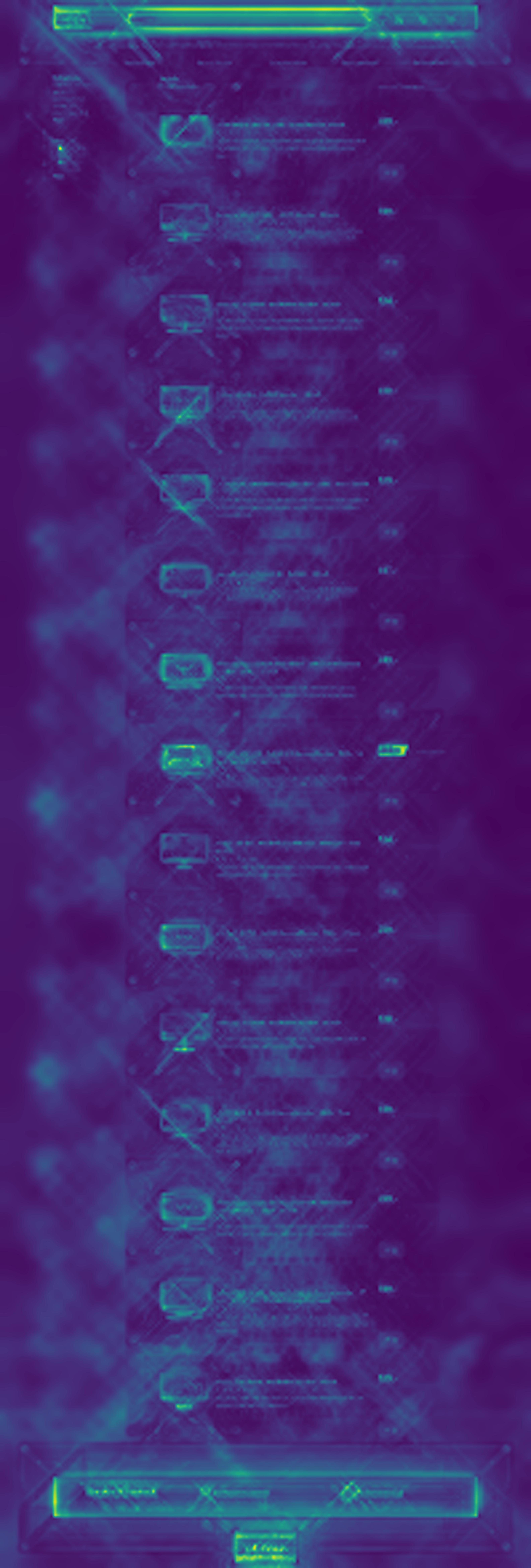}
    \label{fig:visual-sal-maps-monitor-list-itti}
    }
    \subfloat[]{%
    \includegraphics[width=0.26\columnwidth, height=5.2cm]{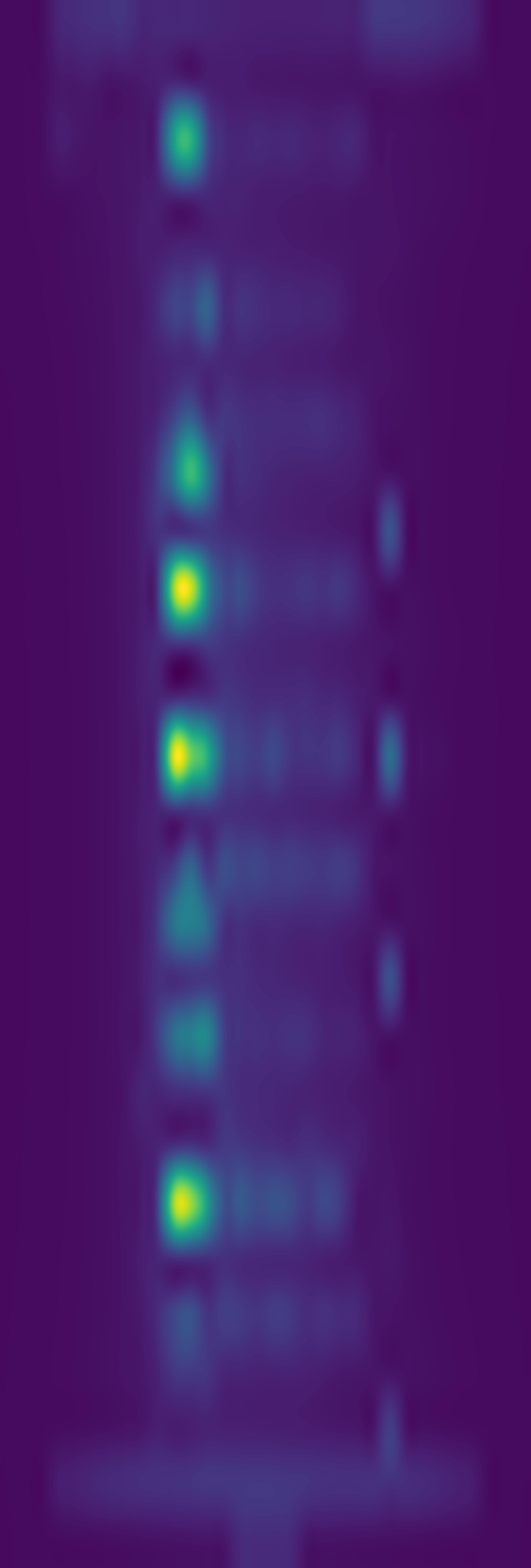}
    \label{fig:visual-sal-maps-monitor-list-gbvs}
    }
    
    \subfloat[]{%
    \includegraphics[width=0.26\columnwidth, height=5.2cm]{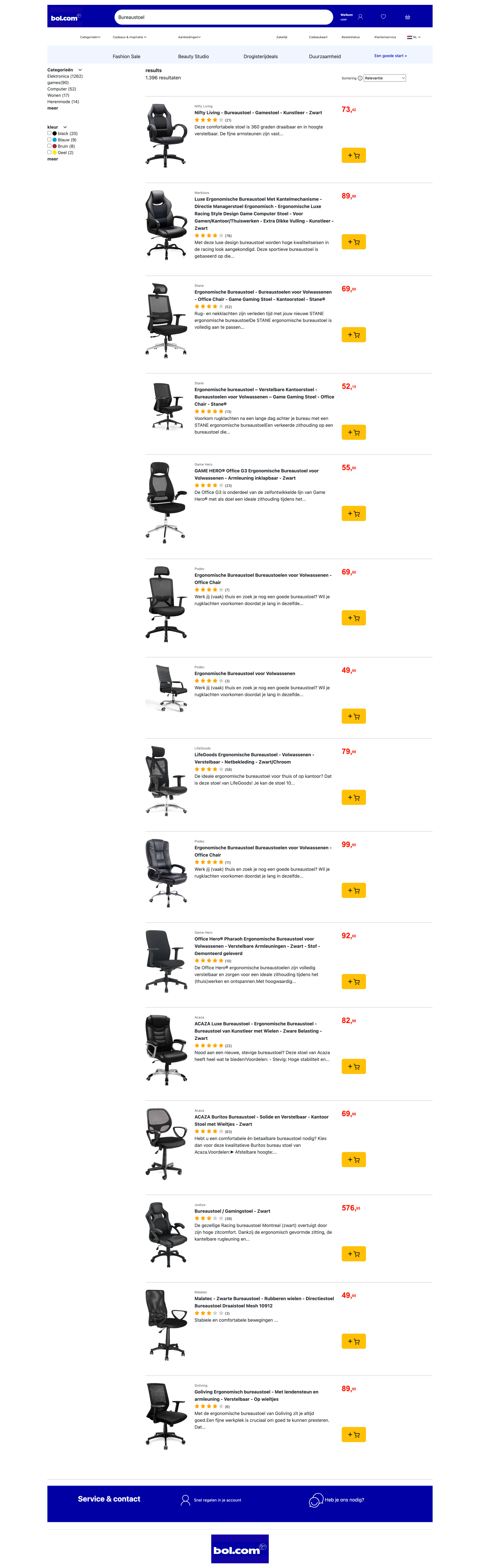}
    \label{fig:visual-sal-maps-chair-list}
    }
    \subfloat[]{%
    \includegraphics[width=0.26\columnwidth, height=5.2cm]{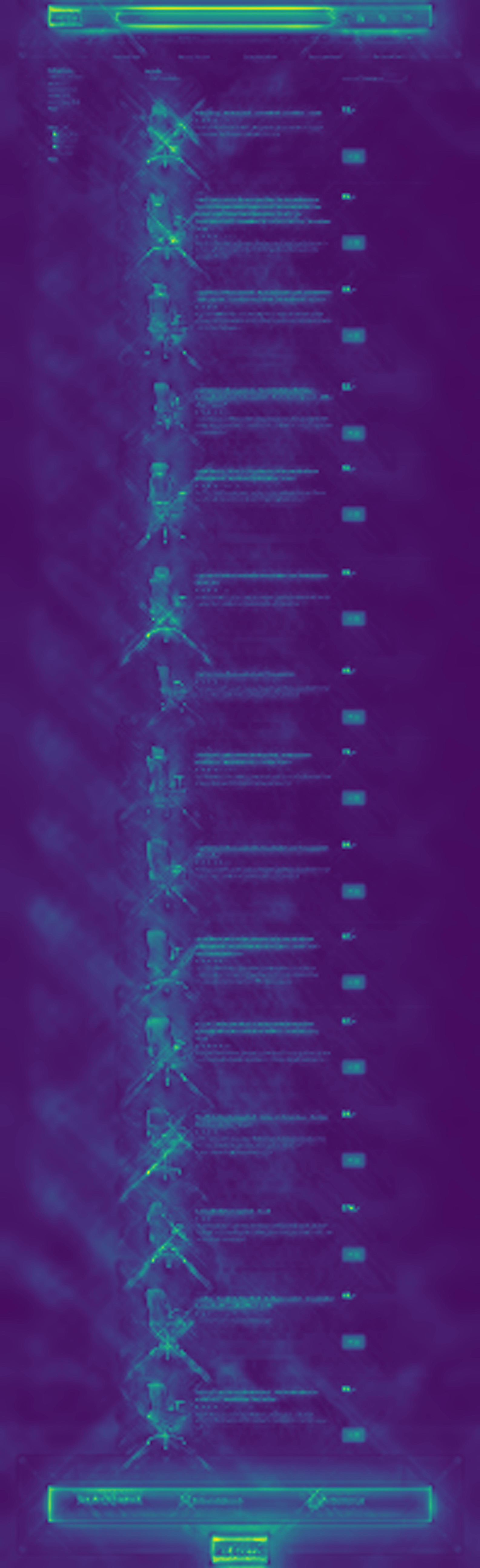}
    \label{fig:visual-sal-maps-chair-itti}
    }
    \subfloat[]{%
    \includegraphics[width=0.26\columnwidth, height=5.2cm]{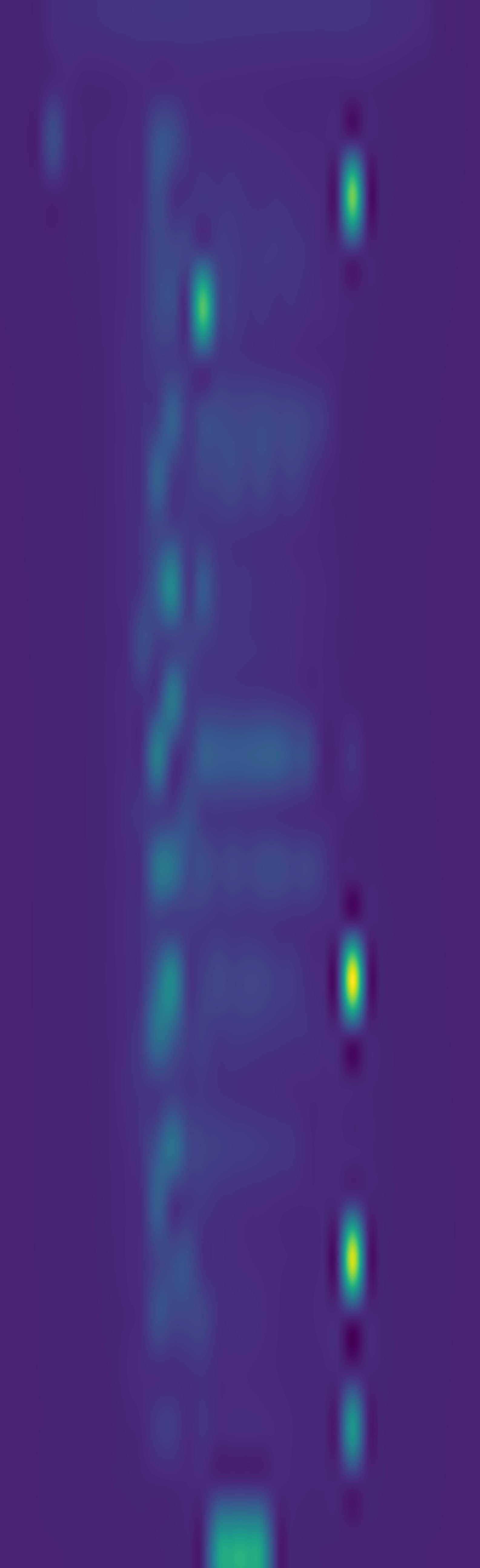}
    \label{fig:visual-sal-maps-chair-list-gbvs}
    }
    \caption{Comparative visualization of actual product lists and predicted visual saliency maps for three distinct product lists:
    (a), (d) and (g) show the original product list for mobile phones with an outlier image at position 3, monitors with an outlier discount tag at position 8, and of fice chairs with an outlier price at position 13, respectively. (b), (e) and (h) show the corresponding visual saliency maps using the \ac{GBVS}, while (c), (f), and (i) show the maps generated by the \itti model.}
    \label{fig:visual-sal-maps-list}
\end{figure}
For the first list, the \ac{GBVS} map highlights a very intense spot near the top where the outlier image is located (see Figure~\ref{fig:visual-sal-maps-mobile-list-gbvs}). This suggests that the model is highly responsive to the visual characteristics of the outlier, potentially due to its unique color scheme, size, and contrast compared to surrounding items. On the other hand, the \itti map shows a more evenly distributed pattern of saliency across the product list (see Figure~\ref{fig:visual-sal-maps-mobile-list-itti}), with less intense focus on any single point. However, there is still a noticeable emphasis on the area around the image outlier, but less pronounced than in the \ac{GBVS} model.

In the second list, the \ac{GBVS} map shows several highlighted areas, but there is a particularly intense focus on the upper part of the list, where the initial items are located (see Figure~\ref{fig:visual-sal-maps-monitor-list-gbvs}). The model does not distinctly highlight the middle part where the outlier discount tag is at position 8, suggesting that while the \ac{GBVS} model is sensitive to certain visual cues, it may not consistently emphasize elements like tags unless they are accompanied by other strong visual contrasts.
While the \itti map displays attention points scattered more evenly across the entire list, there are visible highlights around the middle section, closer to where the outlier discount tag is (see Figure~\ref{fig:visual-sal-maps-monitor-list-itti}). The highlight around the outlier discount tag in the middle of the page might not specifically show the outlier; however, we observe that the \itti map could better detect this local difference in the map. This can be due to its algorithmic sensitivity to a broader range of visual features beyond mere contrast, such as layout structure.

Lastly, in the list with the price outlier, the \ac{GBVS} map shows a few distinct areas of high saliency, with notable intensity at the bottom of the list, near where the outlier is located (see Figure~\ref{fig:visual-sal-maps-chair-list-gbvs}). This suggests that the \ac{GBVS} model is effective in identifying significant deviations in price among this list, while the \itti model fails to detect this pattern (see Figure~\ref{fig:visual-sal-maps-chair-itti}).

In conclusion, our observations suggest that the \ac{GBVS} model is particularly suited for detecting distinct visual anomalies within a cluttered visual field, while the \itti  model offers insights into general visual attention patterns across a product list, by highlighting the more visible parts of the image without focusing on global differences.
The \ac{GBVS} model focuses on global visual features, which means it responds differently based on the complexity and variety of elements in an image. This behavior helps explain why the \ac{GBVS} model fails to detect the bright red tag in Figure~\ref{fig:visual-sal-maps-monitor-list}, while it effectively highlights the price outlier in Figure~\ref{fig:visual-sal-maps-chair-list}. In the chair list, the uniform dark colors of the product images make even small variations in price patterns more noticeable to the model. Conversely, the diverse colors among the monitor images might distract the model, causing it to overlook the red tag despite its visual prominence.

While the \itti and \ac{GBVS} models provide valuable insights into the potential attention-grabbing properties of various product features, they are based on theoretical constructs and may not fully capture real-world user behavior. To address this gap, we now turn to empirical validation through eye-tracking experiments.

\subsubsection{Eye-tracking Task I}
In this section we present our observations on how users examine a regular product result page without any outliers. 

\paragraph{Engagement metrics} 

\begin{figure}[t]
    \centering
    \subfloat[]{\includegraphics[width=0.355\columnwidth]{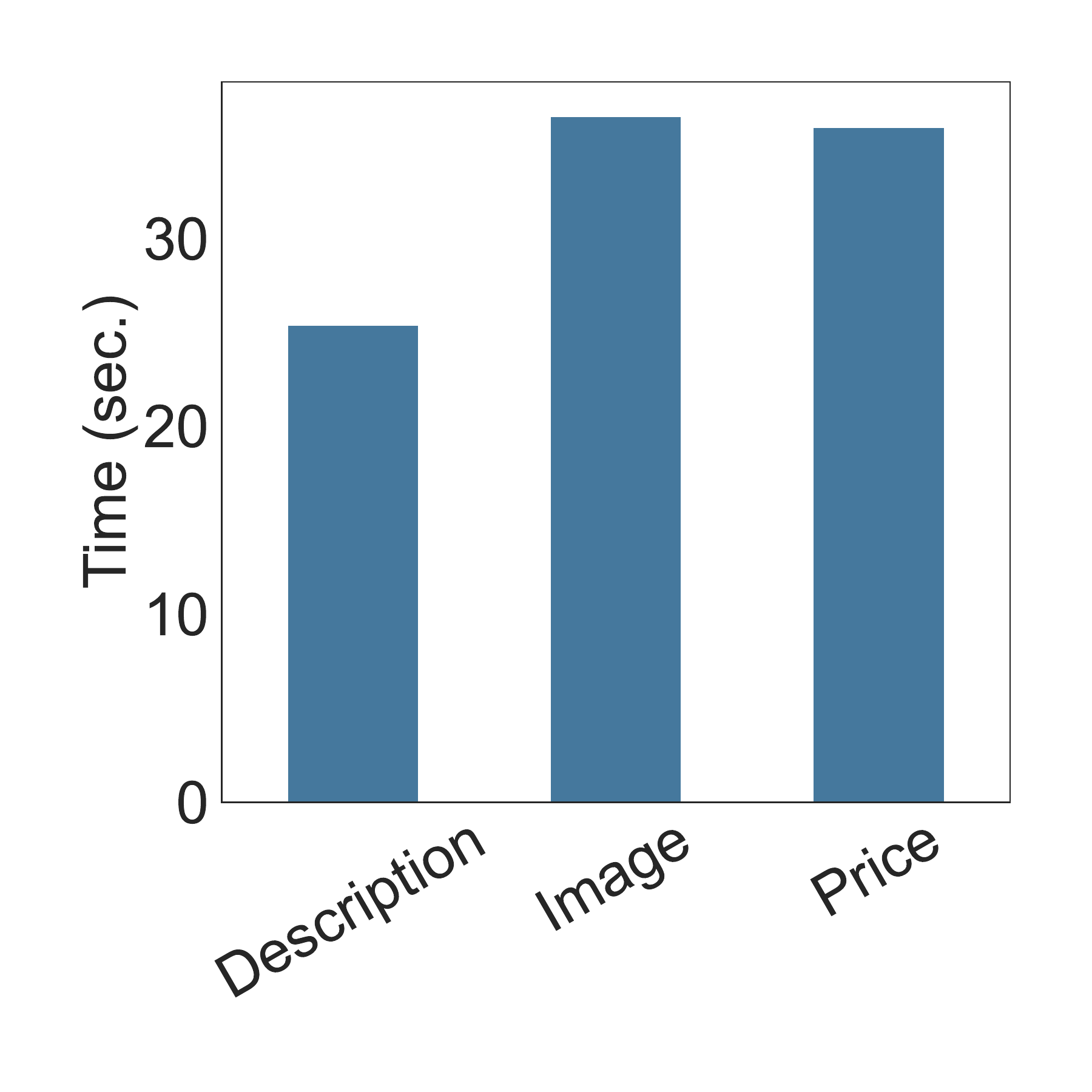}
    \label{fig:barplot-task1-ttff}}
    \subfloat[]{\includegraphics[width=0.355\columnwidth]{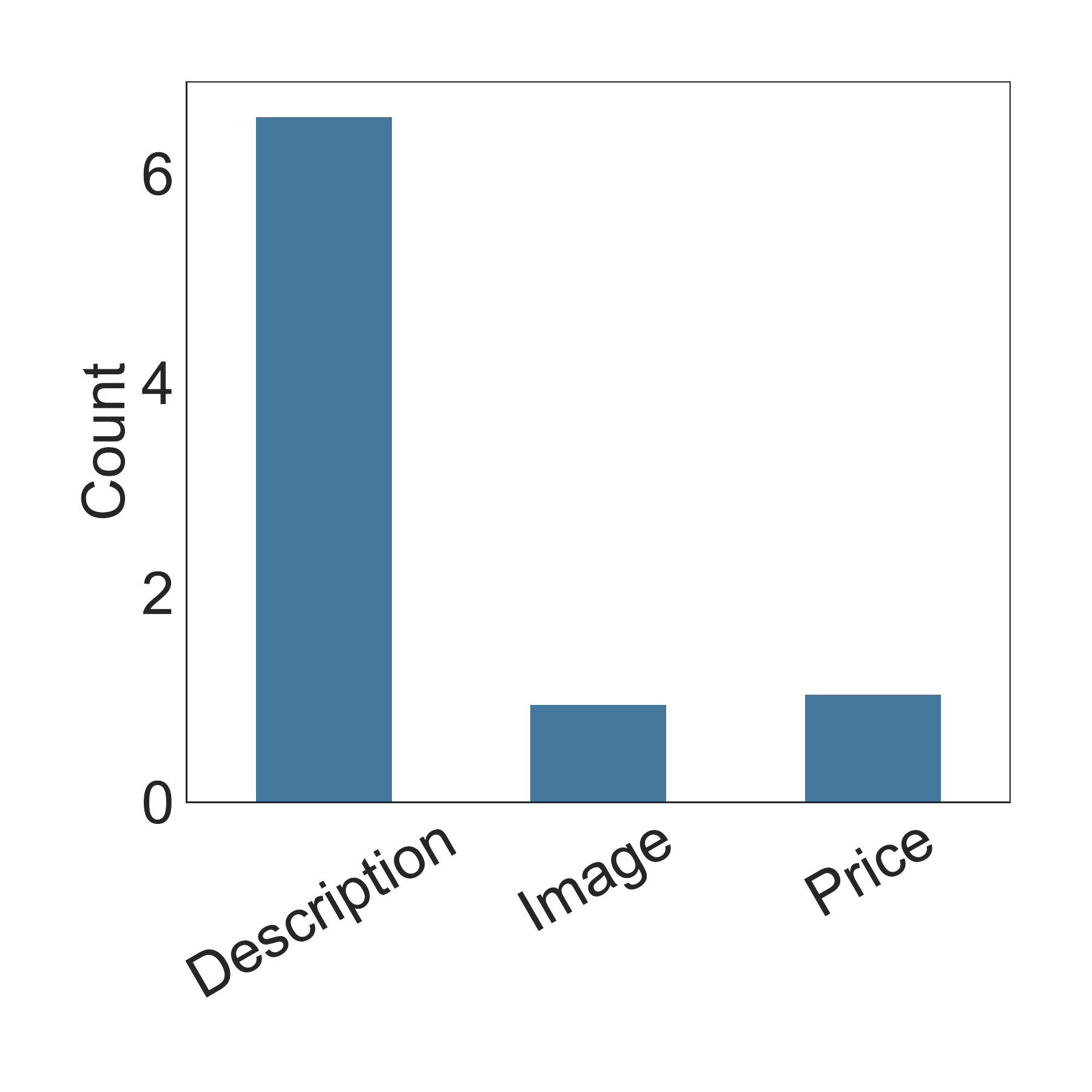}
    \label{fig:barplot-task1-fixation-count}}
    \subfloat[]{\includegraphics[width=0.355\columnwidth]{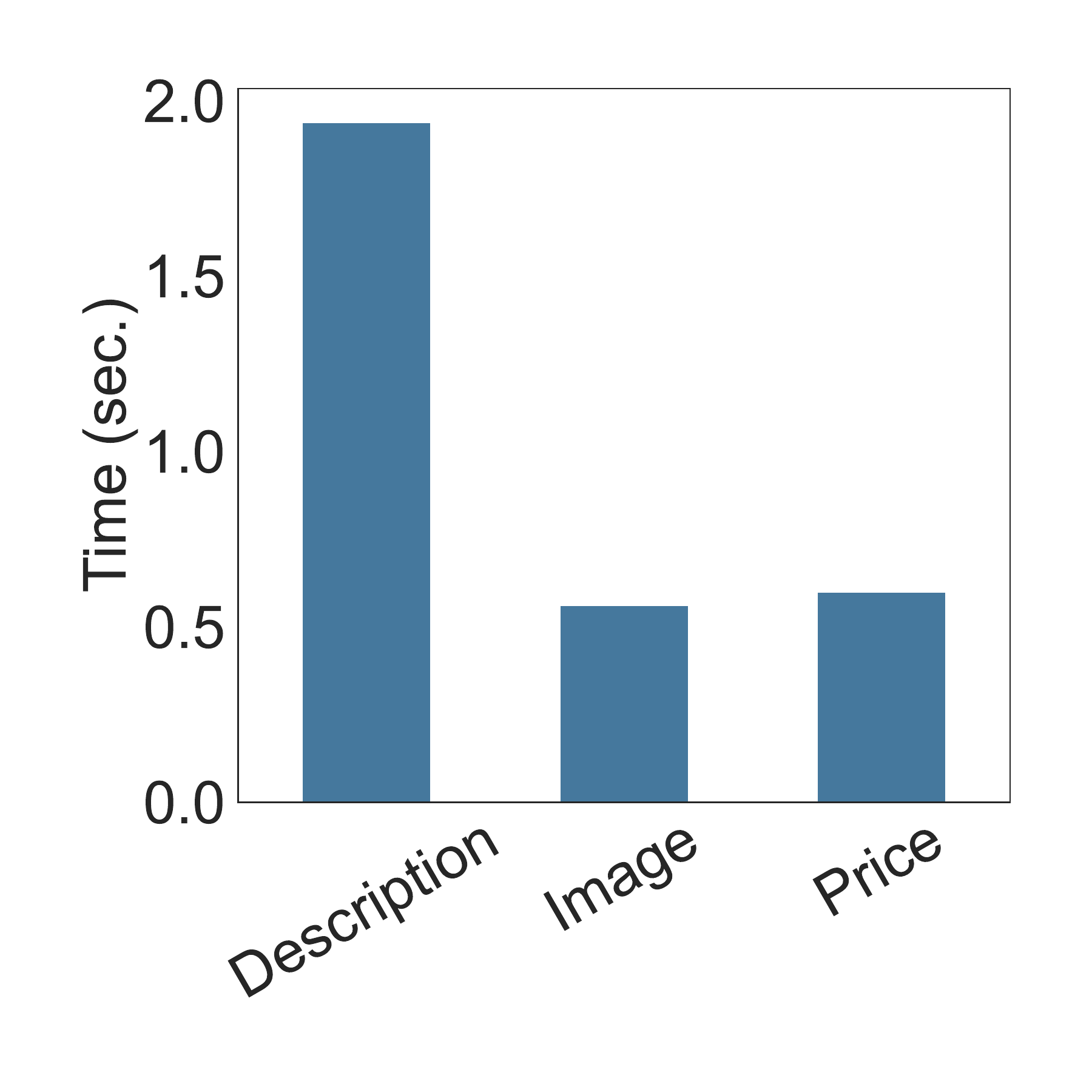}\label{fig:barplot-task1-time-spent}}
    \caption{(a) \ac{TTFF} , (b) average fixation count , and (c) average time spent  by product feature category.}   
    \label{fig:barplot-task1-engagements}
\end{figure}

Figure~\ref{fig:barplot-task1-ttff} shows that \ac{TTFF} is quite high across all categories, with product description being the fastest (25,393 ms), followed by prices (35,901 ms) and images (36,503 ms).
This might indicate that the participants take a significant amount of time before they fixate on any specific element, starting from the middle of the page, confirming center bias~\citep{tatler2007central, buswell1935people, foulsham2008can}. Based on a Kruskal-Wallis test~\citep{mckight2010kruskal} there are statistically significant differences in the \ac{TTFF} across the three feature categories (p-value $<0.05$); however, the difference between \ac{TTFF} on product description and the two other features is more noticeable which could be due to the complexity of scanning the textual information.\footnote{Keep in mind that we instructed users to carefully examine the products.} 

The average fixation count (see Figure~\ref{fig:barplot-task1-fixation-count}) is also significantly higher (Kruskal-Wallis test, p-value $<0.05$) for product description (6.55 times) compared to price (1.03 times) and image (0.94 times), suggesting that once users engage with detailed text, they tend to revisit or focus on these areas more frequently, potentially reflecting a deeper cognitive processing or evaluation.

Consistent with the fixation count, users spend significantly more time (Kruskal-Wallis test, p-value $<0.05$) on product description (1,939 ms) than on  price (600 ms) or image (561 ms) (see Figure~\ref{fig:barplot-task1-time-spent}). This indicates that detailed textual information holds user attention longer.

\begin{figure}[t]
\includegraphics[width=.45\linewidth]{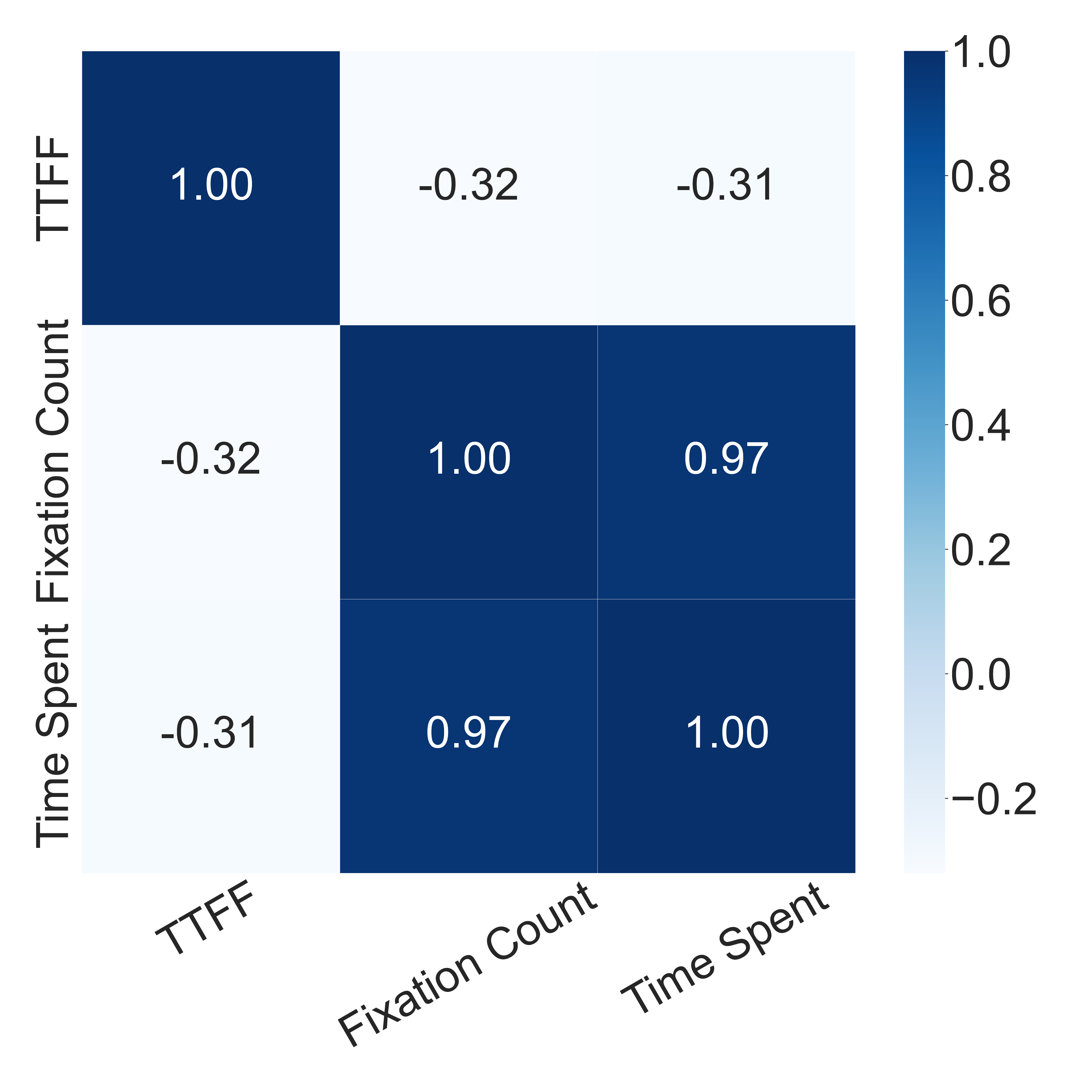}
\caption{Correlation matrix of user engagement metrics based on the eye-tracking experiment.}
\label{fig:heatmap-task1-correlation}
\end{figure}

Moreover, we conducted a correlation analysis between different eye-tracking metrics to see if items that capture attention faster also tend to engage users for longer. Figure~\ref{fig:heatmap-task1-correlation} shows the results. The fixation count and total time spent on different \ac{AOI}s demonstrated a strong correlation (r = 0.80), suggesting that areas that attract more fixations typically engage users for longer duration. This relationship highlights the engagement potential of product description over image or price.

\begin{figure}[h]
    \centering
    \subfloat[]{
    \includegraphics[width=0.55\columnwidth]{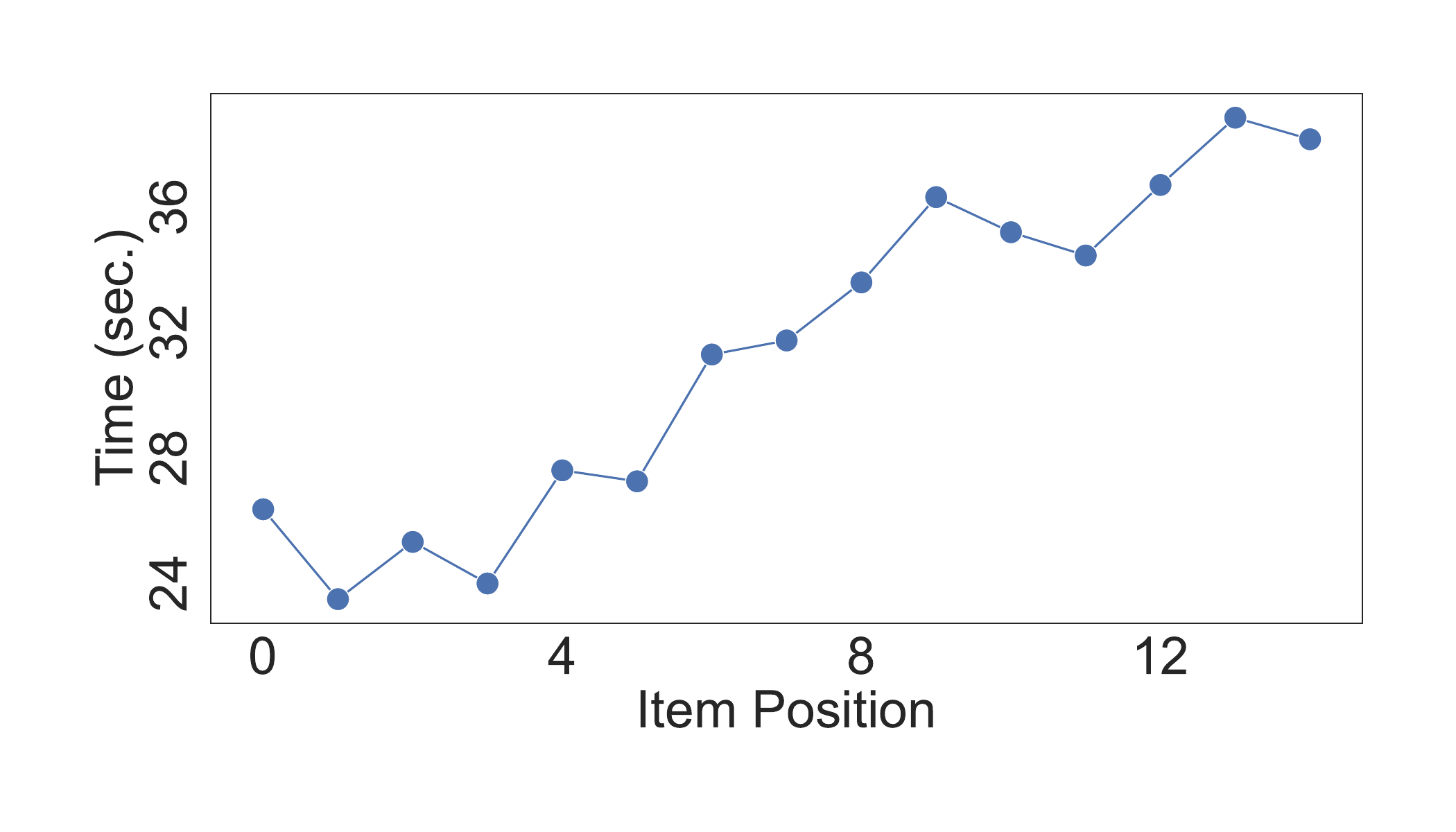}
    \label{fig:lineplot-task1-ttff}}
    
    \subfloat[]{
    \includegraphics[width=0.55\columnwidth]{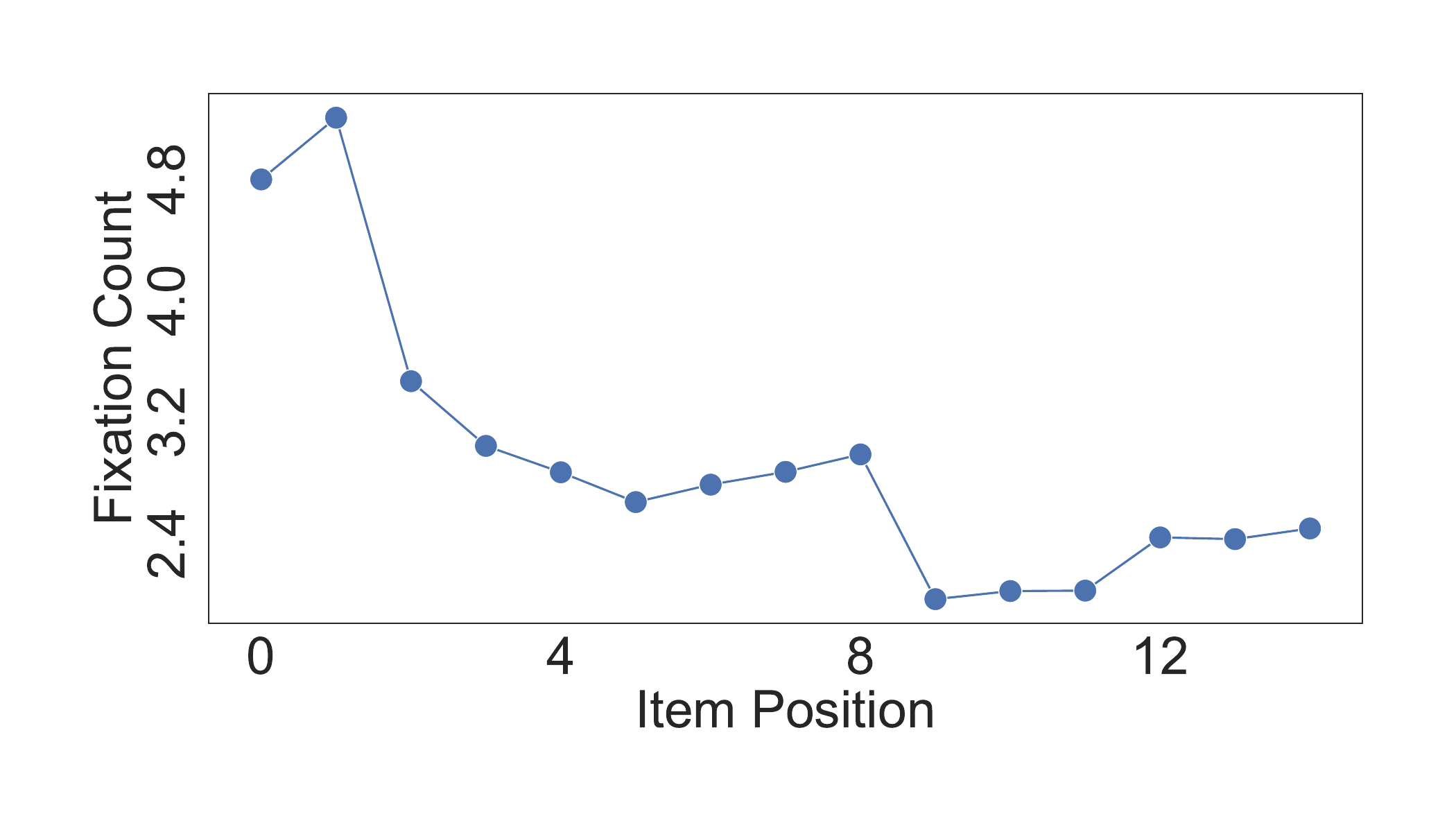}
    \label{fig:lineplot-task1-fixation-count}}
    
    \subfloat[]{
    \includegraphics[width=0.55\columnwidth]{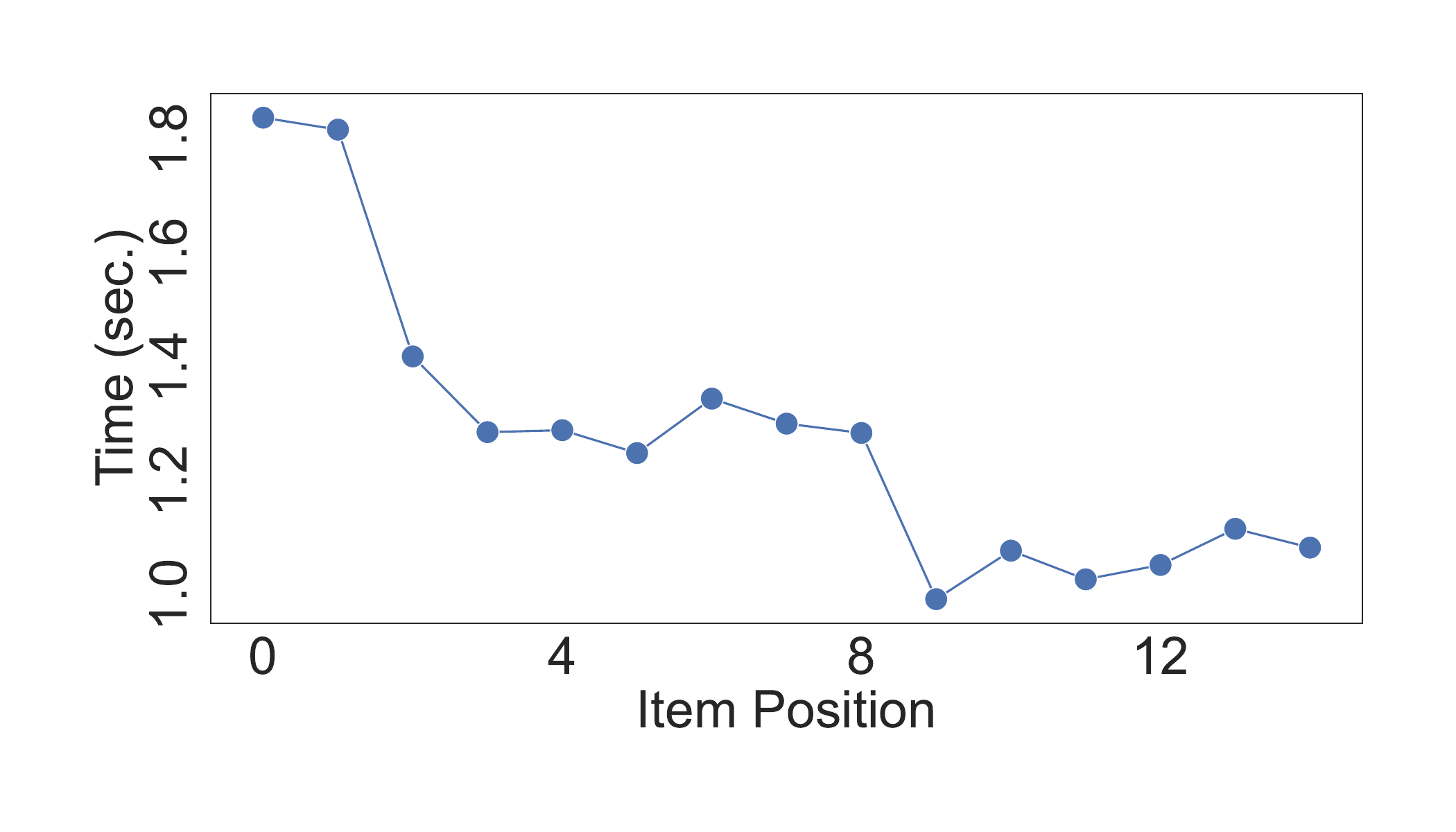}
    \label{fig:lineplot-task1-time-spent}}
    \caption{(a) average \ac{TTFF}, (b) average fixation count and (c) average time spent per position.}   
    \label{fig:lineplot-task1-position-analysis}
\end{figure}

\paragraph{Positional impact analysis}
To understand how the position of an item within a product list affects user interaction, we analyzed \ac{TTFF}, fixation count, and total time spent based on the item's position in the list. Figure~\ref{fig:lineplot-task1-ttff} shows that \ac{TTFF} tends to increase with the position of the item in the list as expected, indicating that users examine the list from top to bottom, and items later in the list take longer to attract the initial fixation. 

The number of fixations generally decreases from the start towards the middle of the list and slightly increases towards the end (see Figure~\ref{fig:lineplot-task1-fixation-count}). This pattern could be influenced by how users scan the page, possibly scanning more quickly through middle items after initially examining the first few items more thoroughly. 

Similar to fixation count, the total time spent also decreases through the list, with the least amount of time spent around the middle of the list (see Figure~\ref{fig:lineplot-task1-time-spent}). This suggests less engagement with items as users move through the list.

In general, these trends indicate that position affects how users interact with items in a product list. Items placed at the beginning of the list are likely to capture attention faster and engage users more than those placed in the middle or bottom of the list.

\subsubsection{Eye-tracking Task II}
To answer~\ref{rq:eye-tracking} we investigate how outlier product features impact the visibility of products in search result lists. This section reports on the findings related to how different features, when presented as outliers, affect visual attention across different list positions. We answer the following sub questions to address~\ref{rq:eye-tracking}:
\begin{enumerate}[label=RQ2-\arabic*]
    \item \label{rq:taskII-1} How do outlier features in product lists influence the initial user attention?
    \item \label{rq:taskII-2} What is the impact of outlier features on user engagement?
\end{enumerate}

\paragraph{\ref{rq:taskII-1}}
We examined how the presence of outlier features in product lists affects initial user attention. Specifically, we focus on~\ac{TTFF} to understand whether products with outlier features attract attention faster. 

Previous work shows that both outliers and their close neighbors attract user attention faster~\citep{sarvi2021understanding}; therefore,
to answer \ref{rq:taskII-1}, we compared~\ac{TTFF} for products with outlier features and their immediate neighbors against more distant neighbors.  Immediate neighbors are defined as the products immediately preceding and following the outlier, while distant neighbors are those either before or after the immediate neighbors. 
We analyzed this setup across the outlier features (image, price, discount tag) and different product positions (3, 8, 13).

\begin{figure}[t]
    \centering
    \subfloat[\label{fig:barplot-task2-RQ1-image}]{
    \includegraphics[width=0.3\columnwidth]{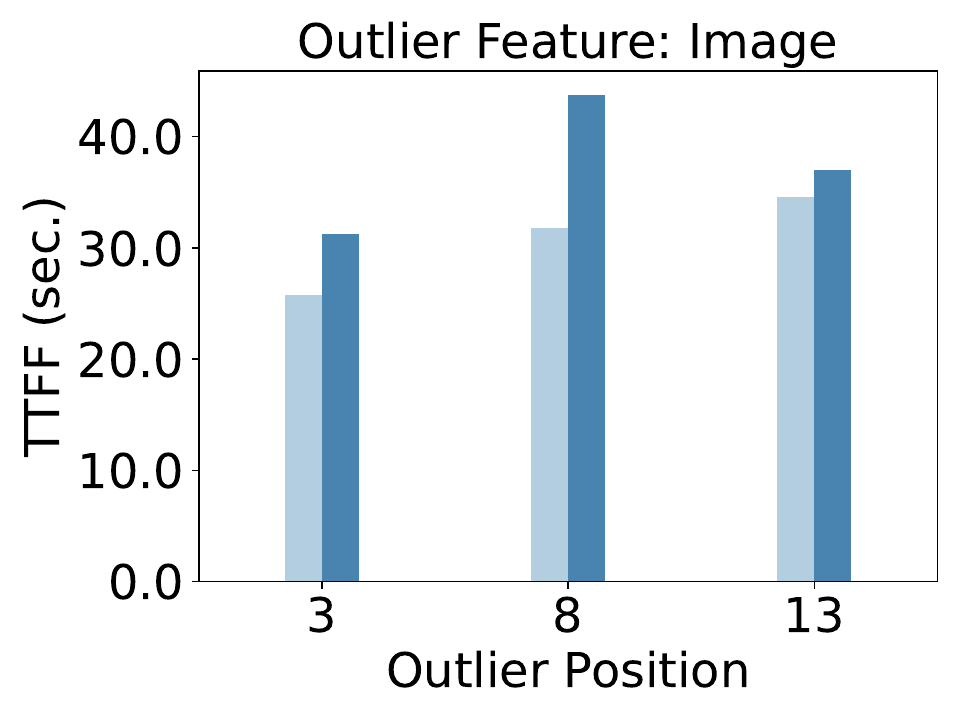}}
    \subfloat[\label{fig:barplot-task2-RQ1-price}]{\includegraphics[width=0.3\columnwidth]{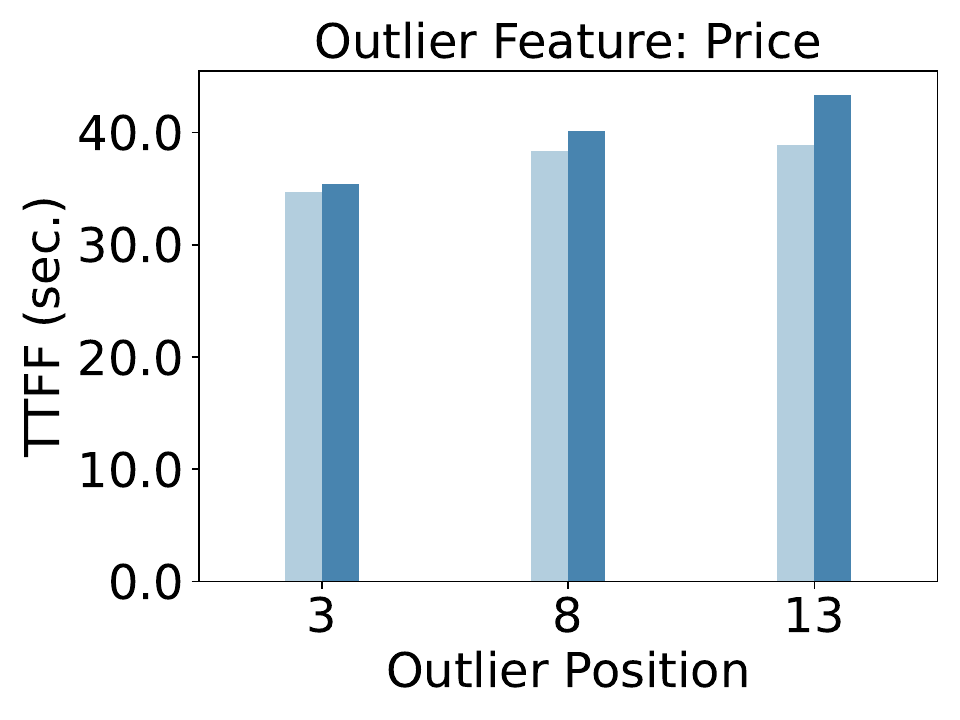}}
    \subfloat[\label{fig:barplot-task2-RQ1-tag}]{\includegraphics[width=0.3\columnwidth]{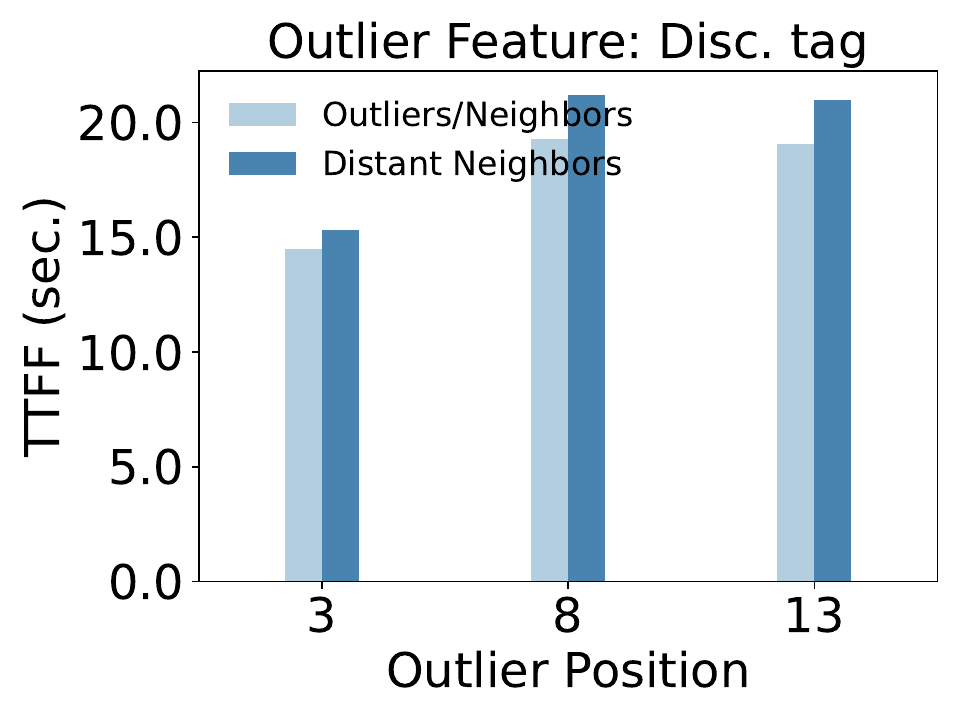}}
    \caption{\ac{TTFF} for (a) image, (b) price, and (c) discount tag outliers per position, showing the comparison between outliers and their immediate neighbors versus their distant neighbors.}   
    \label{fig:barplot-task2-RQ1}
\end{figure}

\begin{figure}[h]
\includegraphics[width=1\columnwidth]{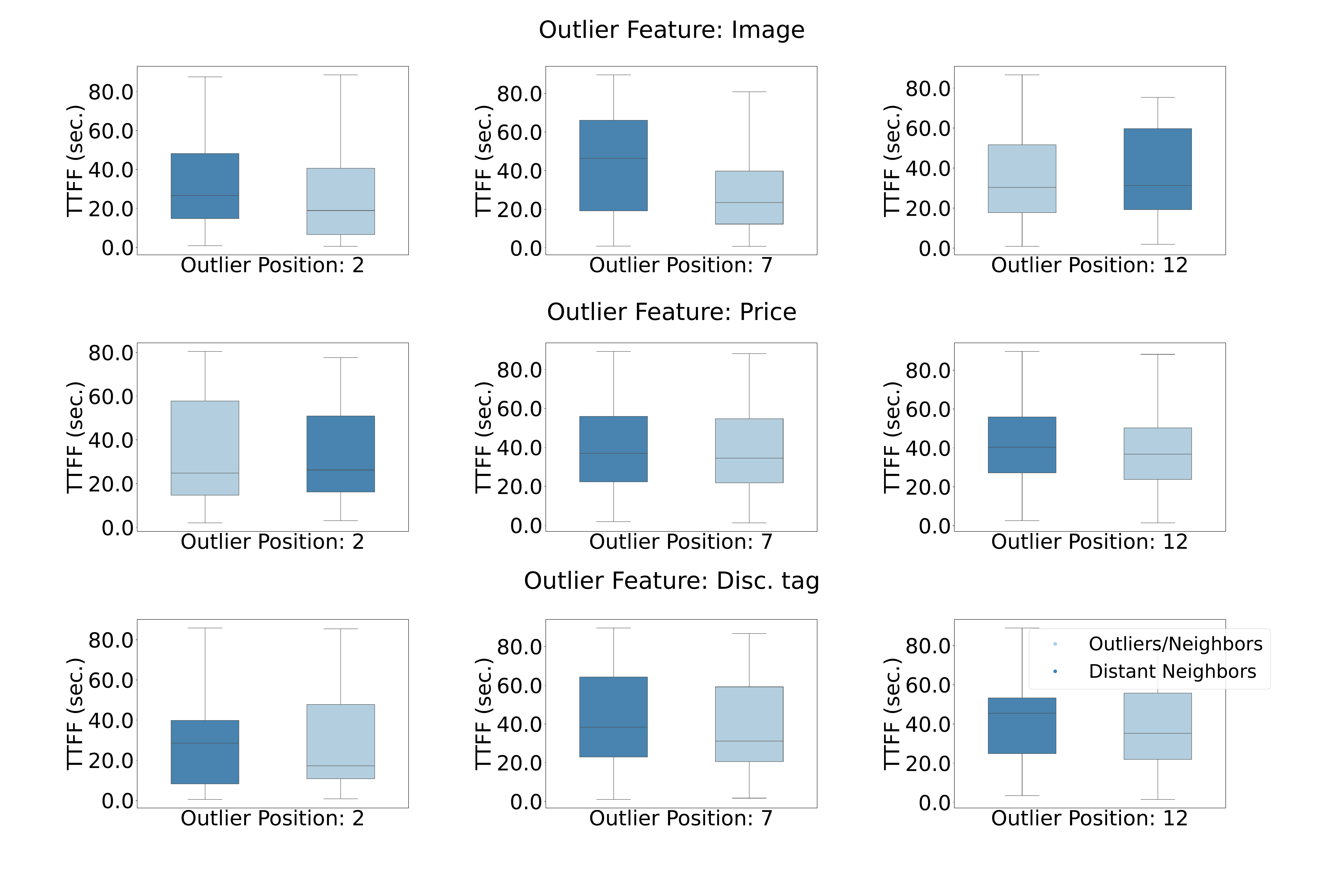}
\caption{Distribution and range of~\ac{TTFF} values across different positions and outlier features.}
\label{fig:boxplot-task2-RQ1}
\end{figure}

For image outliers, Figure~\ref{fig:barplot-task2-RQ1-image} shows that outliers and their  immediate neighbors at all positions generally exhibit a lower mean~\ac{TTFF} compared to distant neighbors, indicating quicker attention capture ($25.76s$ vs. $31.24s$ for position 3, $31.77s$ vs. $43.71s$ for position 8 and $34.57s$ vs. $37.03s$ for position 13). 
Similar trends are observed with price and discount tag features (see Figure~\ref{fig:barplot-task2-RQ1-price} and Figure~\ref{fig:barplot-task2-RQ1-tag}), where outliers and their immediate neighbors consistently show lower~\ac{TTFF} compared to distant neighbors, although differences were less pronounced compared to image features. A Kruskal-Wallis test shows statistically significant differences between~\ac{TTFF} of the two groups among different outlier features and positions with p-value $<0.05$. Figure~\ref{fig:boxplot-task2-RQ1} details the distribution and range of~\ac{TTFF} values.

Our results confirm that outlier features not only draw attention more quickly but also potentially increase the exposure of their adjacent items. This effect is consistent across different types of features and various positions within the list.

\begin{figure}[h]
\includegraphics[width=1\columnwidth]{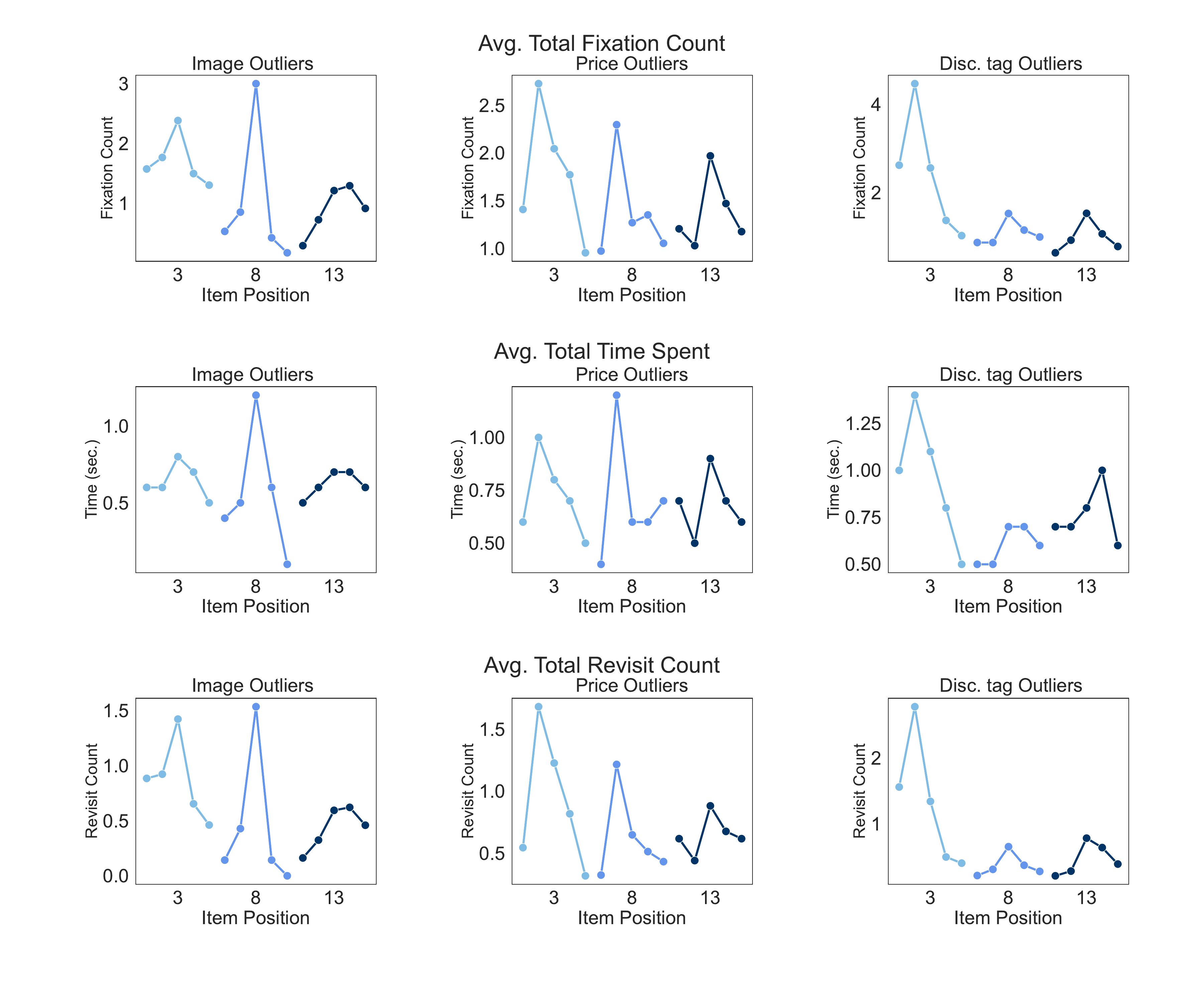}
\caption{Average user engagement metrics across all users, calculated for outliers, their immediate neighbors, and distant neighbors, across different positions and outlier features.}
\label{fig:task2-RQ2-user-engagement}
\end{figure}

\paragraph{\ref{rq:taskII-2}}
To address~\ref{rq:taskII-2}, we analyze how outlier features influence user engagement by examining metrics such as total fixation count, time spent and revisit counts. These metrics help us understand not just the initial attention (as explored in~\ref{rq:taskII-1}) but also the sustained interest and engagement. Figure~\ref{fig:task2-RQ2-user-engagement} depicts the average values of these metrics calculated over all users for different features, outlier positions, their immediate and more distant neighbors. 

Starting with the total fixation count, we observe peaks at positions 3, 8, and 13 for image outliers, indicating significant engagement at all positions, but particularly strong at position 8 where the average fixation count reaches $\approx 2.5$. This suggests that image outliers, regardless of their position, tend to draw consistent attention, with a focus on the middle of the list.

Price outliers show similar peaks at these positions with the highest at position 3 ($\approx 2.75$), indicating that price outliers at the start of the list may capture slightly more attention than those later, possibly due to immediate price evaluation when beginning the list browsing.

Discount tag outliers exhibit the highest fixation count at the start of the list (position 3 with a count of about 4), with noticeable decreases thereafter. This highlights that discount tags catch the eye quickly, possibly due to the initial scanning behavior of users.

The two other metrics, time spent and revisit counts, follow the same trends for different variants, emphasizing that outlier items and their immediate neighbors not only capture user attention faster than other items in the list, but also receive more exposure and user engagement.  

\subsubsection{Comparison and upshot}
The \itti and \ac{GBVS} models predict attention hotspots using visual features like color, contrast, and size. 
The \itti model provides a more evenly distributed attention pattern, recognizing areas with notable visual differences without intense focus on a single point.
In contrast, the \ac{GBVS} model highlights prominent anomalies with strong visual contrasts, successfully identifying image, and price outliers due to their unique visual characteristics. 

However, these models fail to consistently detect outliers in the product lists. This limitation can be attributed to the fact that the models only rely on the bottom-up factors, which has two main implications. First, the models see the whole list as one picture. For instance, in our experiments, the diverse colors in the monitor list distracted the model from the red tag (see Figure~\ref{fig:visual-sal-maps-monitor-list-gbvs}), while the uniform dark colors in the chair list made small price variations more noticeable. 

Second, the models ignore the top-down factors such as users' sensitivity to discounts or their tendency to compare specific features like discount tags across products. The models cannot compare any specific features across different products.

Observations from our eye-tracking Task I support the top-down factors at play. Specifically, users engaged more with product descriptions, which, despite being less visually attractive, held significant attention in terms of both \ac{TTFF} (that can also be explained by center bias) and sustained engagement (see Figure~\ref{fig:barplot-task1-engagements}). 
This engagement with detailed information highlights the importance of user intent, something that visual saliency models do not account for.

The empirical results from our second eye-tracking experiment provide a more comprehensive understanding of user attention and engagement in the presence of outliers. The experiments confirmed that outliers capture attention quickly, as evidenced by lower \ac{TTFF} for outliers compared to distant neighbors. Eye-tracking data also revealed that users engaged with these outliers for longer durations, providing insights into sustained engagement that visual saliency models fail to capture.

In conclusion, our observations suggest that visual saliency models are effective in predicting initial visual attraction based on basic visual properties. They are useful for quick, preliminary assessments and designing visually appealing interfaces. However, their reliance on the bottom-up factors limits their applicability in fully understanding user behavior in complex, real-world scenarios. In contrast, eye-tracking experiments, while more resource-intensive, provide comprehensive insights into both initial attention and sustained engagement. They account for top-down factors, capturing real-world user interactions more accurately.


\section{Discussion \& Conclusion}
\label{sec:discussion}

\subsection{Research problem and objectives}
In this study, we explored how different presentational features influence the perception of outliers in e-commerce search results through a two-stage approach.

We designed the initial visual search experiments to explore how features such as price, star rating, and discount tags affect users' ability to identify outliers. These experiments provided initial insights into the immediate observability of these features and the impact of visual complexity on user perception.

Building on these preliminary findings, our subsequent eye-tracking experiments aimed to validate and extend our understanding by observing user behavior in a more realistic simulated e-commerce environment. These experiments measured actual user attention and engagement, providing a more comprehensive view of how outlier features capture and sustain attention in real-world-like scenarios.
We also incorporated visual saliency analysis to predict which product features would naturally attract attention due to their visual properties. This analysis served as a benchmark to compare with empirical eye-tracking data, allowing us to understand the interplay between bottom-up visual factors and top-down cognitive factors in shaping user attention during online shopping.

\subsection{Main findings}
Our initial visual search experiments suggest that visual complexity of a feature affects item outlierness. The visual saliency models confirm this observation by consistently highlighting areas with strong visual contrasts, distinct colors, and complex patterns as attention hotspots. This is expected since these algorithms work based on bottom-up visual factors. Moreover, Figure~\ref{fig:barplot-task2-RQ1} shows that, averaged over all outlier positions, \ac{TTFF} is the lowest for image outliers, followed by discount tags, and the highest for price outliers ($30.70s$, $35.21s$, and $37.3s$, respectively). Although the product lists used in the eye-tracking experiments were different, the overall trend is in line with our observations from visual search experiments. 

Additionally, our observations emphasize that one should be cautious about the limitations of visual saliency models in such contexts. As mentioned, visual saliency models only rely on bottom-up factors, making them naive in that they do not distinguish between separate product features or comparing them against each other, instead they analyze the entire image and highlight areas that stand out based on overall visual complexity. Therefore, in lists with colorful and complex product images, the models might miss an obvious outlier such as a unique discount tag, while detecting a subtle visual difference made by a higher price in a list with uniform dark colors. 

Moreover, our eye-tracking Task I suggests that despite being less visually attractive, product descriptions captured attention more quickly, indicating the importance of the top-down factors and other factors in play like center bias. This is evident by a lower \ac{TTFF} for product descriptions in Figure~\ref{fig:barplot-task1-ttff} followed by more revisits and time spent on these areas, reflecting deeper cognitive engagement (see Figures~\ref{fig:barplot-task1-fixation-count} and~\ref{fig:barplot-task1-time-spent}).

In Task II of our eye-tracking study, we examined how outlier product features influence visibility in search result lists. The results indicated that outliers and their immediate neighbors attracted attention faster (in terms of \ac{TTFF}) and engaged users for longer durations (in terms of fixation count and time spent) compared to distant items. This effect was consistent across different outlier features (image, price, discount tag) and various positions within the list. 

Overall, our findings from visual search, visual saliency models, and eye-tracking experiments emphasize the dual role of visual and cognitive factors in shaping user attention. Visual saliency models effectively predict initial visual attraction based on bottom-up factors, while eye-tracking data provides comprehensive insights into sustained engagement driven by the top-down factors. These insights can inform the design of more effective and engaging e-commerce interfaces by optimizing the presentation of key product features to capture and maintain user attention.

\subsection{Limitations and future work}
Although our work provides insights into how different features contribute to the outlierness of an item, it does not directly estimate outlier feature exposure, which is influenced by various user behaviors and platform-specific algorithms not covered in this study. We focused on common presentational features, but item presentation can vary significantly among e-commerce platforms. Additionally, we limited our study to list view presentations, while findings might differ in grid views or other layouts. Lastly, our online eye-tracking data collection introduces potential noise from technical and environmental factors that are out of our control.

Future work should focus on quantifying and generalizing the impact of different visual attributes on item outlierness to understand their broader implications. Additionally, developing models that predict areas of attention by integrating both top-down and bottom-up factors could be highly beneficial. Using eye-tracking data as training inputs for visual saliency maps tailored to the e-commerce domain could enhance these models' accuracy and applicability.

\section*{Resources}
To facilitate reproducibility of the work in this paper, all code and parameters are shared at \url{https://github.com/arezooSarvi/outlier-visual-saliency}.

\begin{acks}
This research was supported by 
Ahold Delhaize,
the Hybrid Intelligence Center, a 10-year program funded by the Dutch Ministry of Education, Culture and Science through the Netherlands Organisation for Scientific Research, \url{https://hybrid-intelligence-centre.nl}, project nr.\ 024.004.022,
project LESSEN with project number NWA.1389.20.183 of the research program NWA ORC 2020/21, which is (partly) financed by the Dutch Research Council (NWO),
project ROBUST with project number KICH3.LTP.20.006, which is (partly) financed by the Dutch Research Council (NWO), DPG Media, RTL, and the Dutch Ministry of Economic Affairs and Climate Policy (EZK) under the program LTP KIC 2020-2023,
and the FINDHR (Fairness and Intersectional Non-Discrimination in Human Recommendation) project that received funding from the European Union's Horizon Europe research and innovation program under grant agreement No 101070212,

All content represents the opinion of the authors, which is not necessarily shared or endorsed by their respective employers and/or sponsors.

We would like to thank Shubha Guha for her valuable contributions to the preparation and execution of the experiments in this work.
\end{acks}

\appendix
\section{Appendix}

\begin{figure}[h]
    \subfloat[]{%
    \includegraphics[width=0.4\columnwidth, height=4.5cm]{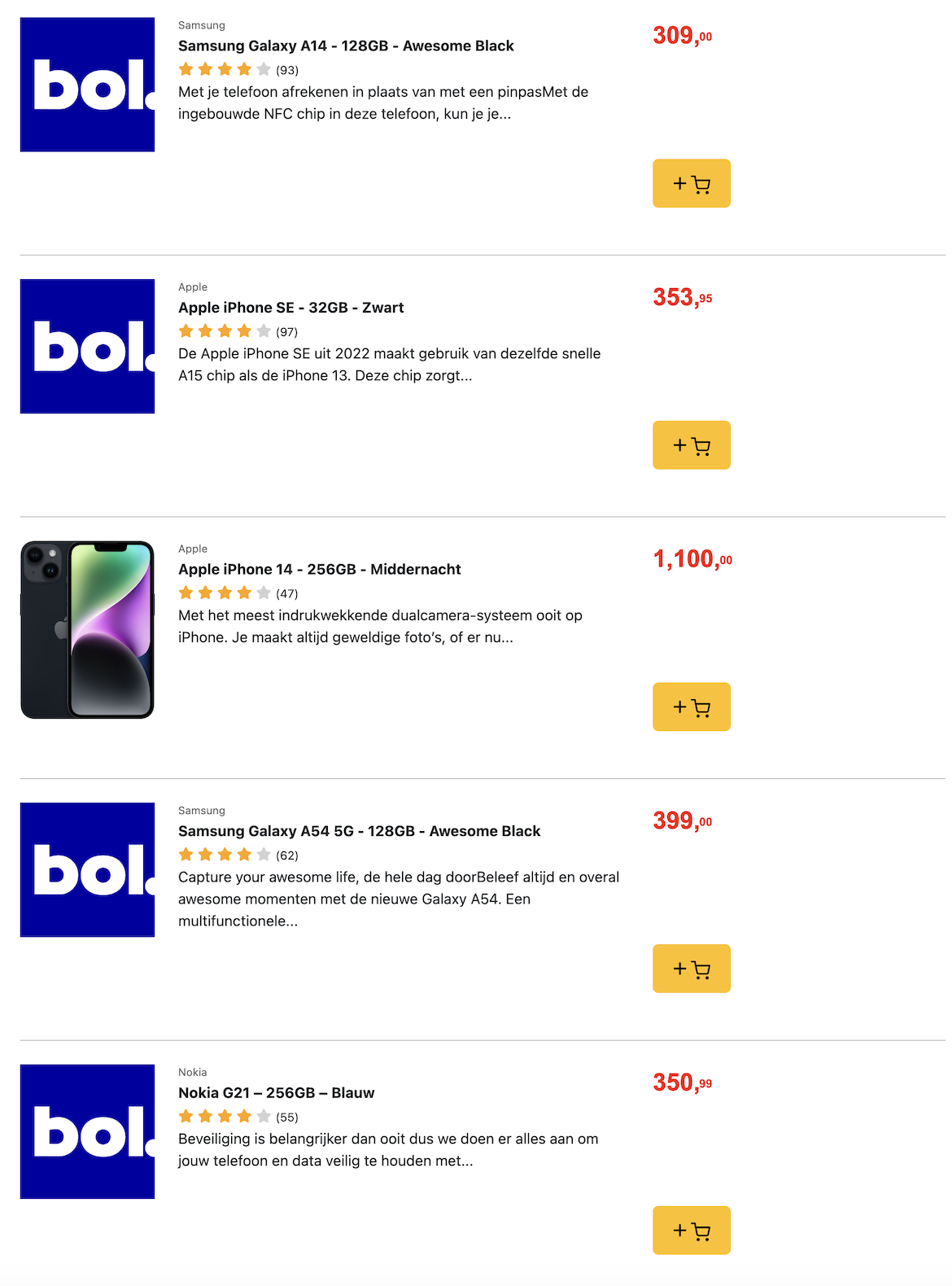}
    }
    \hfill
    \subfloat[]{%
    \includegraphics[width=0.25\columnwidth, height=4.5cm]{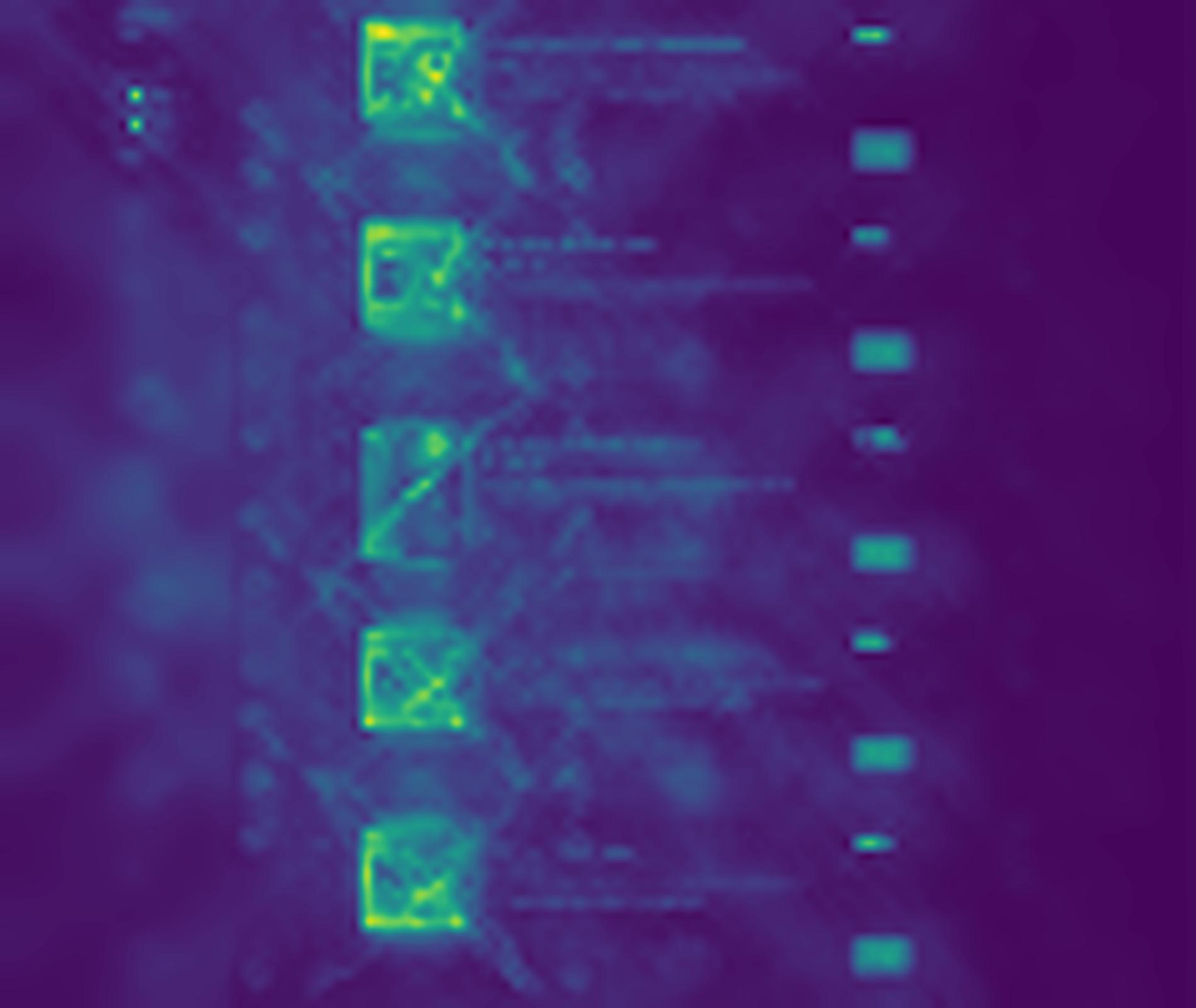}
    }
    \hfill
    \subfloat[]{%
    \includegraphics[width=0.25\columnwidth, height=4.5cm]{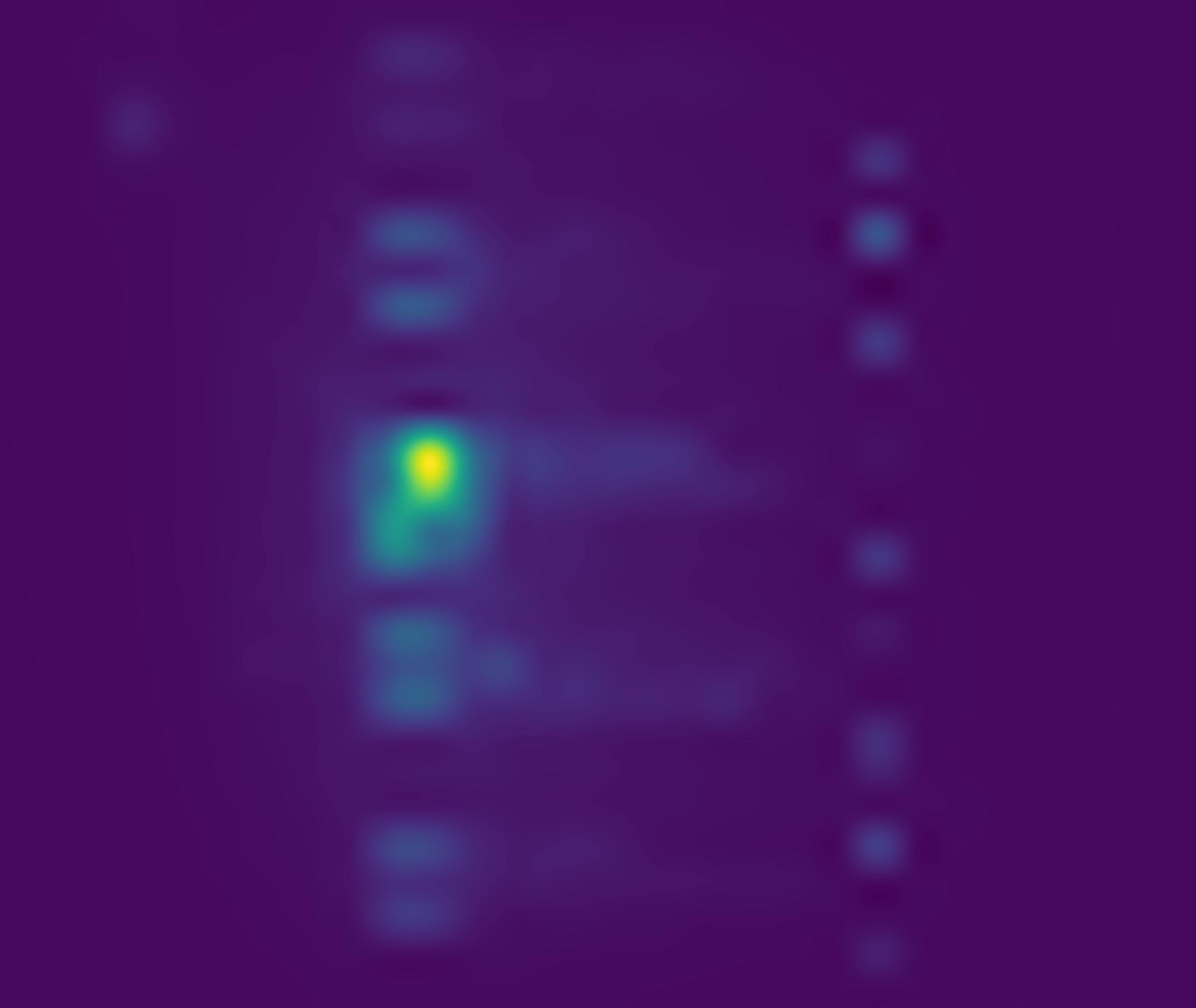}
    }

    \subfloat[]{%
        \includegraphics[width=0.4\columnwidth, height=4.5cm]{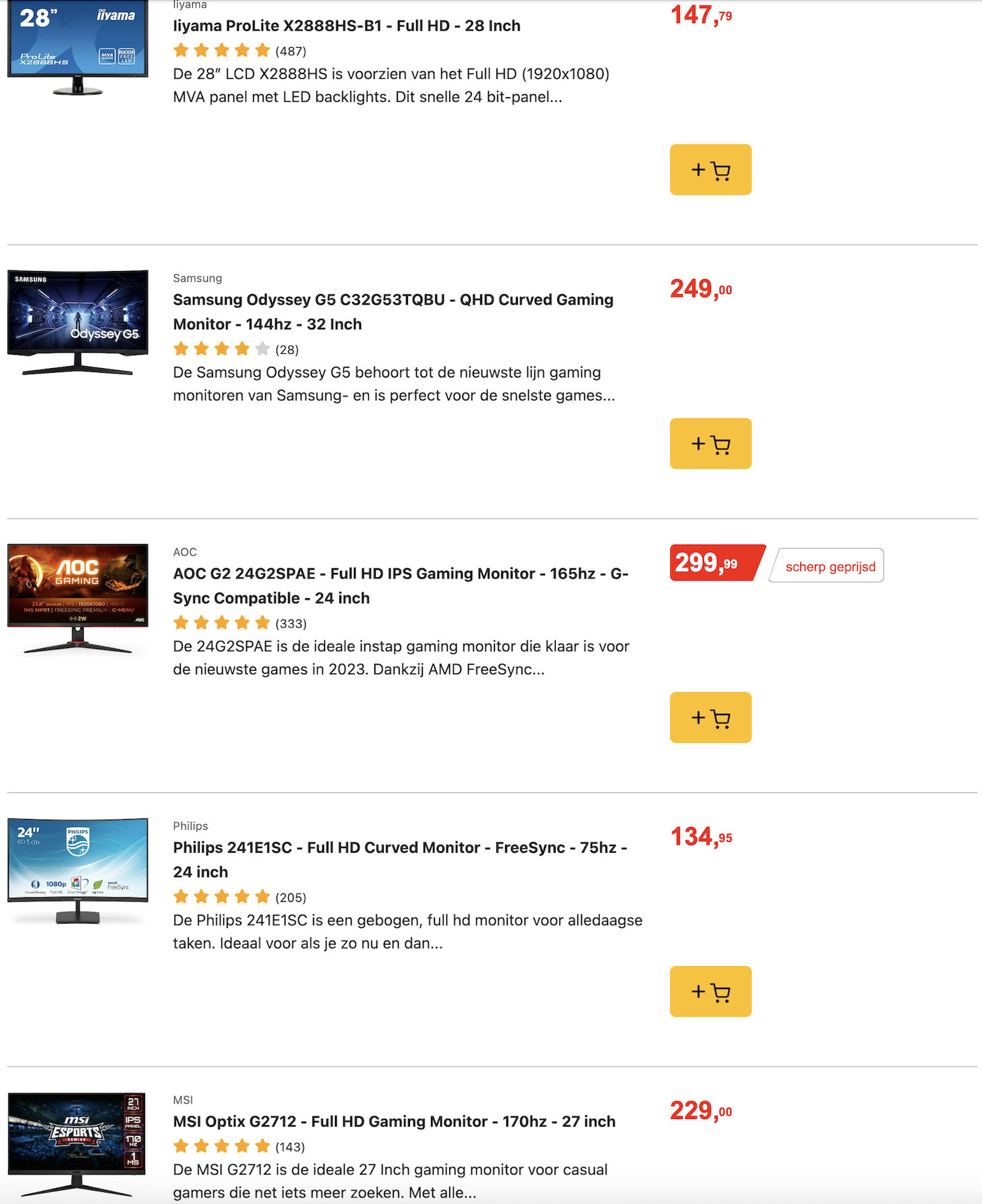}
    }
    \hfill
    \subfloat[]{%
    \includegraphics[width=0.25\columnwidth, height=4.5cm]{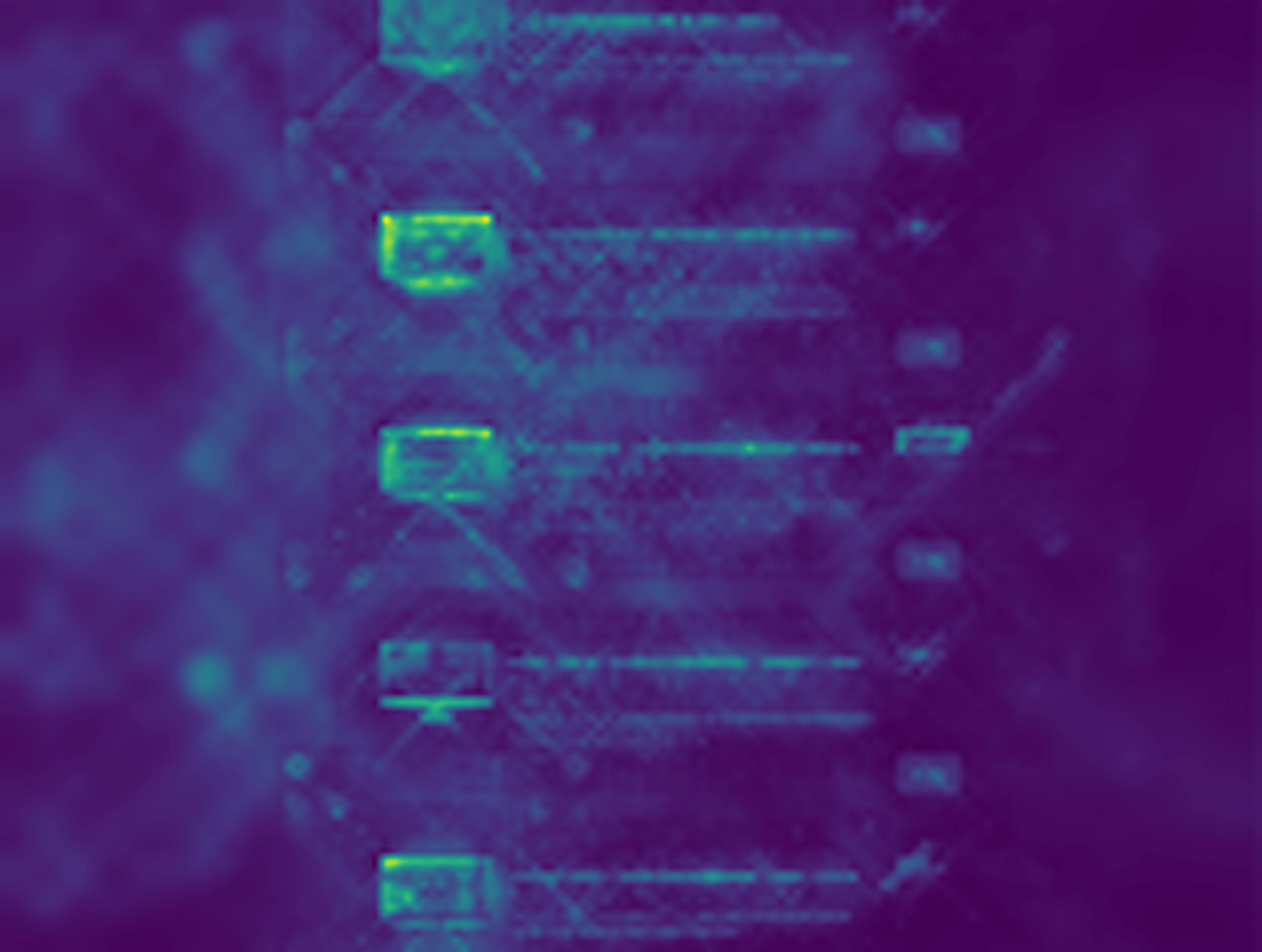}
    }
    \hfill
    \subfloat[]{%
    \includegraphics[width=0.25\columnwidth, height=4.5cm]{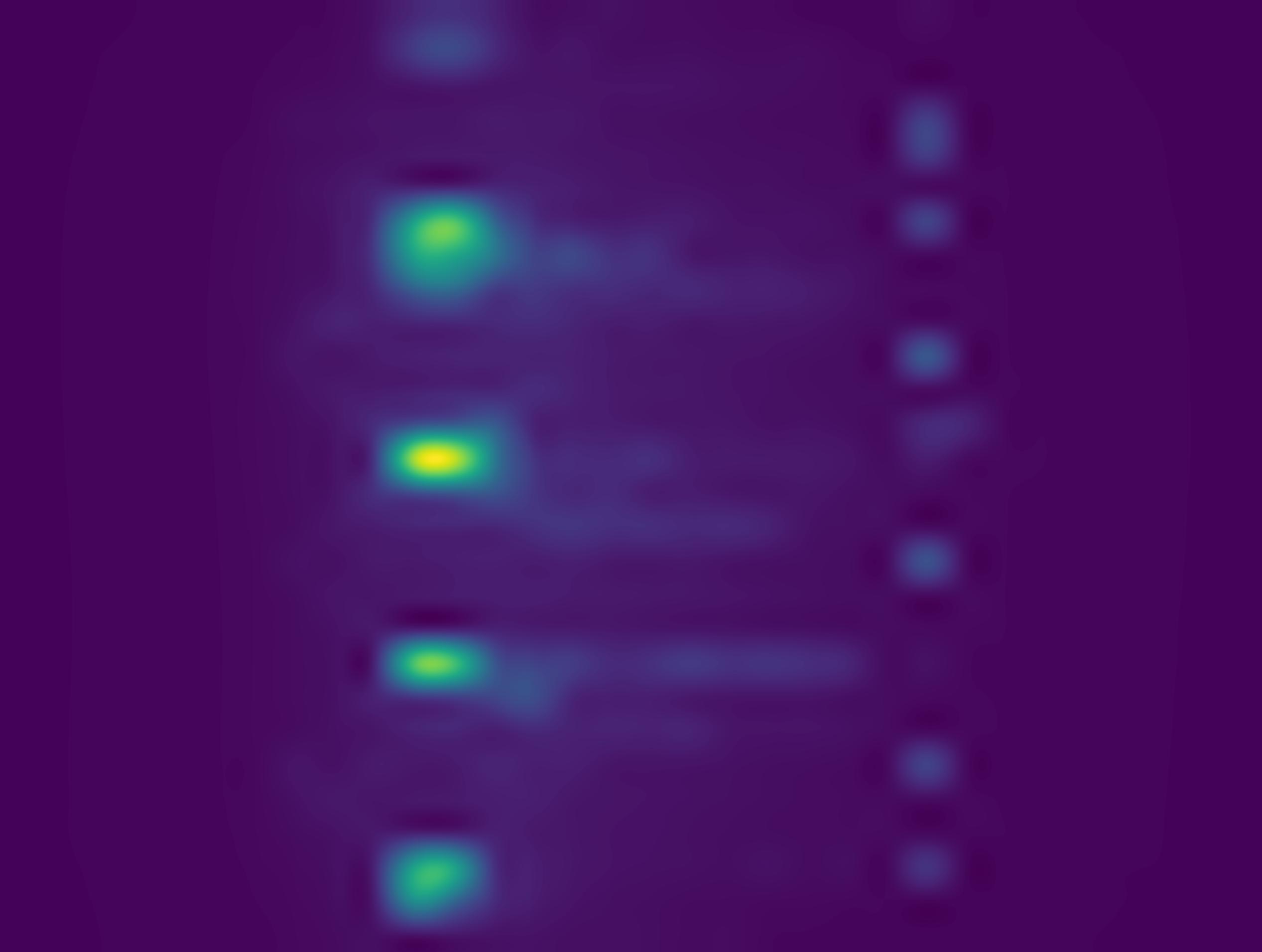}
    }
    
    \subfloat[]{%
        \includegraphics[width=0.4\columnwidth, height=4.5cm]{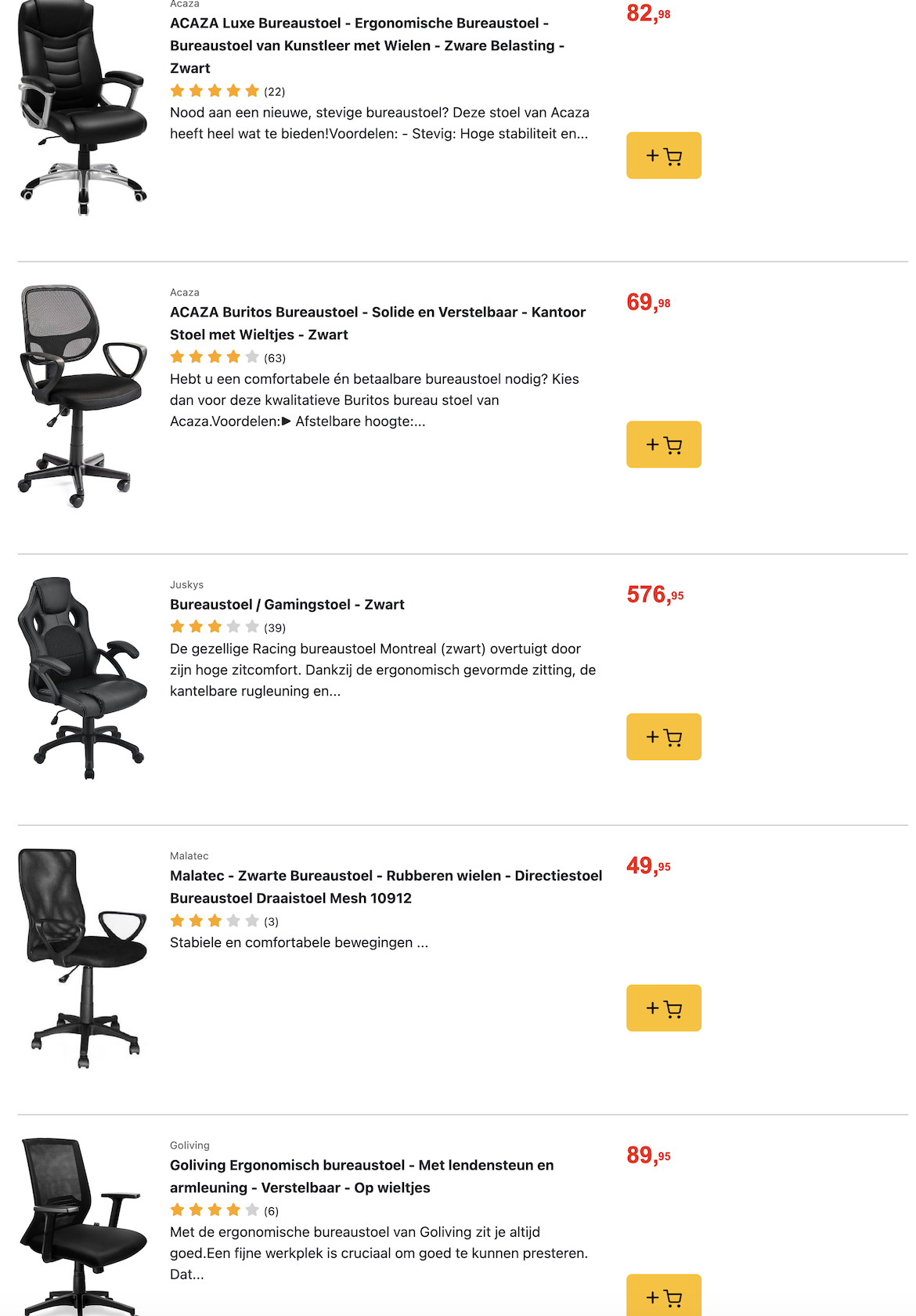}
    }
    \hfill
    \subfloat[]{%
        \includegraphics[width=0.25\columnwidth, height=4.5cm]{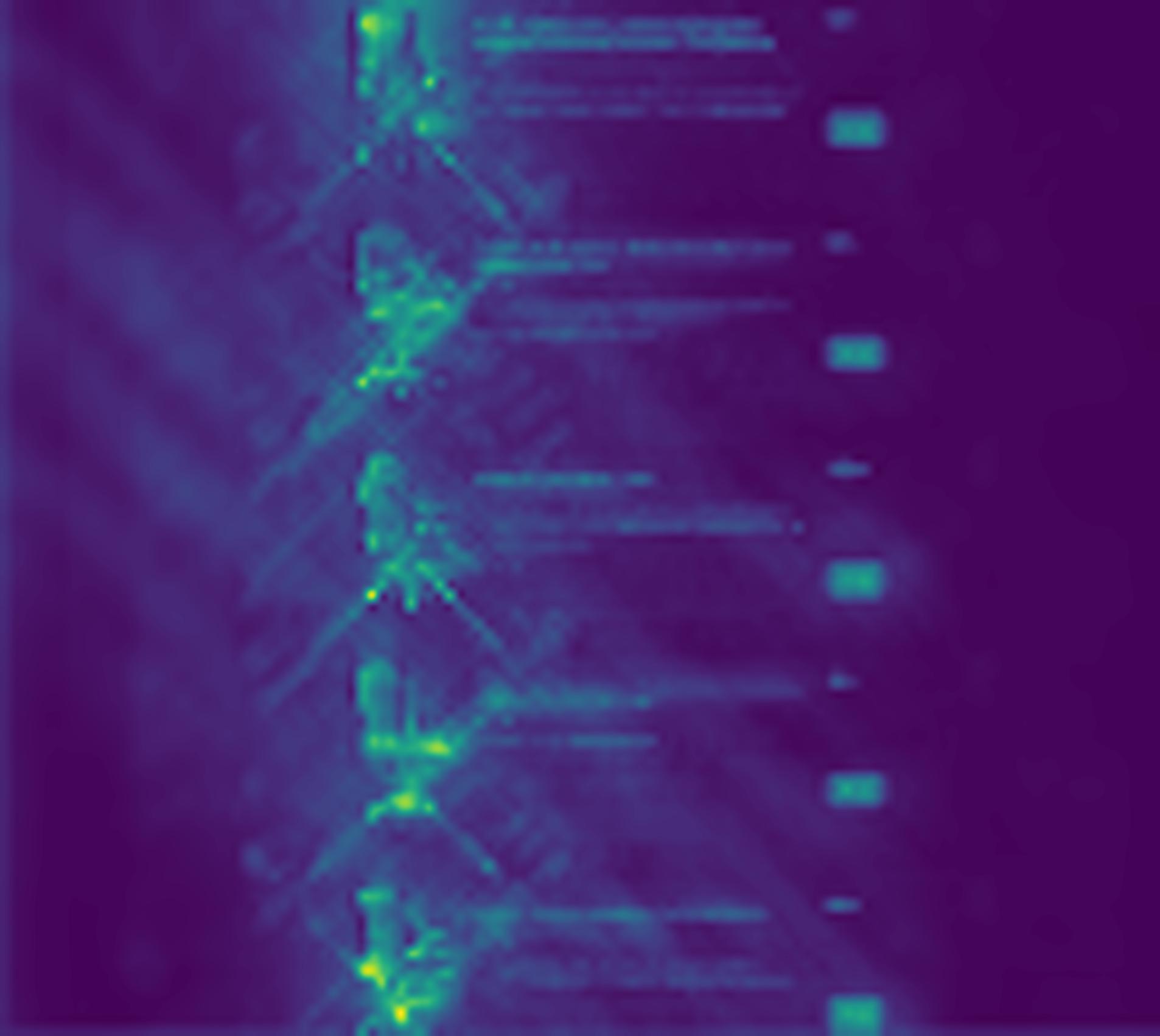}
    }
    \hfill
    \subfloat[]{%
    \includegraphics[width=0.25\columnwidth, height=4.5cm]{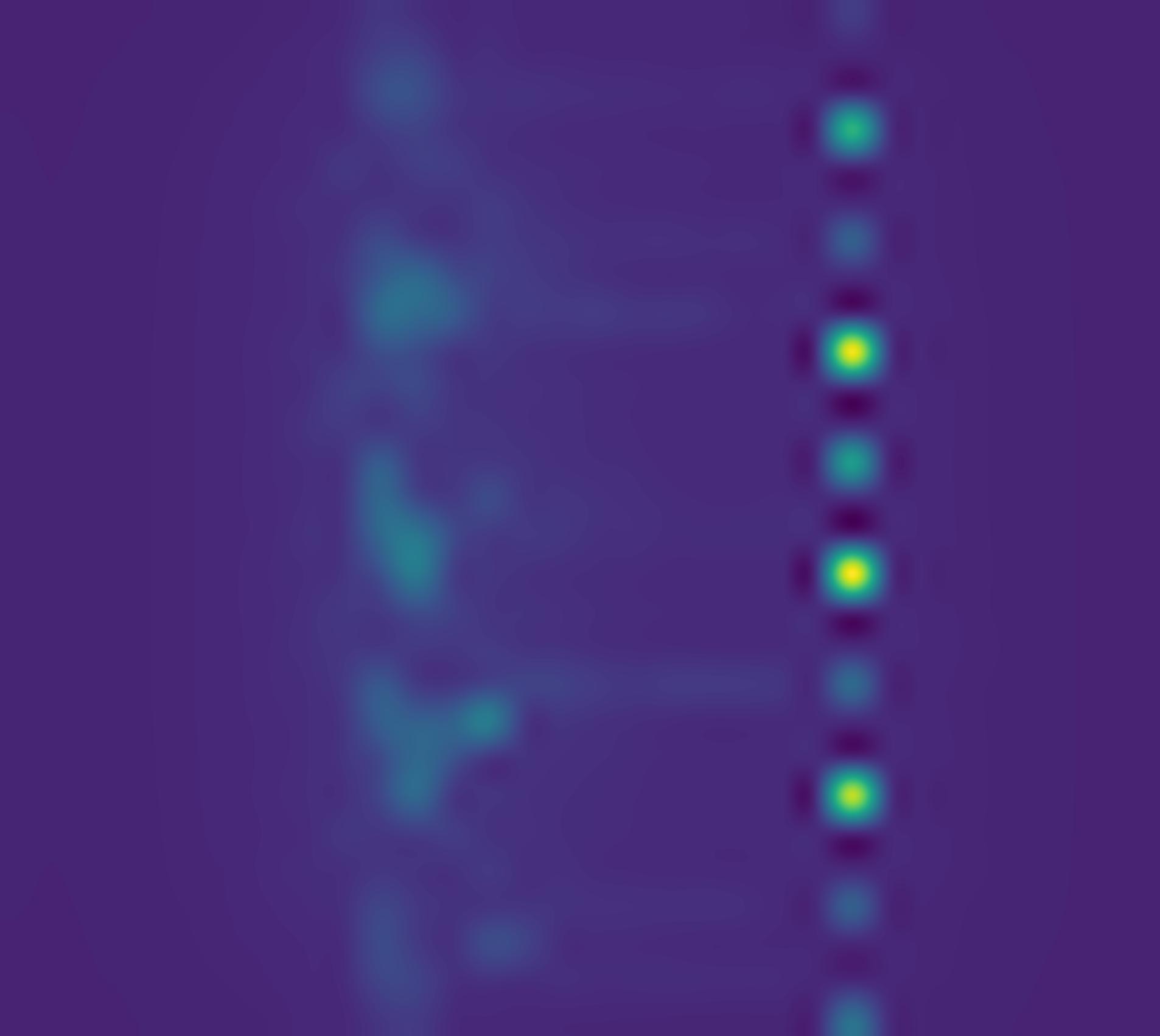}
    }
    \caption{Comparative visualization of focused product lists and predicted visual saliency maps:
    (a), (d) and (g) show the original product list with only the outlier item and its close neighbors, for mobile phones at position 3, for monitors at position 8 and for office chairs at position 13, respectively. (b), (e) and (h) show the corresponding visual saliency maps using the \ac{GBVS}, while (c), (f) and (i) show the maps generated by the \itti model.}
    \label{appendix-fig:visual-sal-maps-list}
\end{figure}

\subsection{Task instructions}
In the following, we provide the instructions that were shared with participants on the Prolific platform for our experiments:

\vspace{0.5em}
\noindent
In this task, you'll review some product lists on an online shopping website (Bol.com). You'll explore a specific category, like smartphones, shoes, or backpacks, as if you're planning to make a purchase.

We require \textbf{access to your webcam} to record your \textbf{eye movements} as you review the product lists. Please be assured that we will only record eye movements on the product list pages, and your personal or private data will not be recorded or accessed in any way.

\vspace{0.5em}
\noindent
Please carefully examine all the products on the page. \textbf{We'll ask questions about your observations afterward}, such as:
\begin{itemize}
    \item Describe what you saw on the list briefly, e.g., price range, item types.
    \item Note if you observed a specific gender focus or prominent colors.
    \item List any brands you noticed.
    \item Recommend one item for purchase and explain why.
    \item Mention anything that caught your attention.
\end{itemize}

Your answers can be in English or Dutch.
\textbf{IMPORTANT}: You \textbf{MUST} answer the post-task questions about the content of the lists accurately, otherwise we \textbf{CANNOT ACCEPT} your submission.

\bibliographystyle{ACM-Reference-Format}
\balance
\bibliography{references}

\end{document}